\documentclass[a4paper,fleqn]{cas-dc}  % fleqn = left-aligned equations

\usepackage[numbers]{natbib}   % numbered citations
\usepackage{amsmath,amssymb,amsfonts}
\usepackage{graphicx}
\usepackage{tikz}
\usepackage{tcolorbox}
\usepackage{pdflscape}
\usepackage{rotating}
\usepackage{longtable}
\usepackage{multirow}
\usepackage{enumitem}
\usepackage{fontawesome}
\usepackage{tablefootnote}
\usepackage{colortbl}
\usepackage[export]{adjustbox}
\usepackage{framed}
\usepackage{svg}
\usepackage{subfig}
\usepackage{placeins}
\usepackage{url}            % simple URL typesetting
\usepackage{hyperref}
\usepackage{xurl}
\usepackage{array}
\usepackage{float}
\usepackage{hyperref}
\usepackage{balance}
\usepackage{tabularx}
\usepackage{booktabs}
\usepackage{ragged2e}
\usepackage{longtable}
\usepackage[table]{xcolor}  % enables \rowcolor in tables
\definecolor{lightgrayrow}{gray}{0.92}
\newcolumntype{L}[1]{>{\RaggedRight\arraybackslash}p{#1}}

\newcolumntype{Y}{>{\RaggedRight\arraybackslash}X}
\begin{document}
\begin{sloppypar}

\let\WriteBookmarks\relax
\def\floatpagepagefraction{1}
\def\textpagefraction{.001}

\shorttitle{LLM-Based Multi-Agent Systems for Code Generation}
\shortauthors{Rasheed et al.}

\title[mode=title]{LLM-Based Multi-Agent Systems for Code Generation: A Multi-Vocal Literature Review\tnotemark[1]}

% ---- Authors ----
\author[tampere]{Zeeshan Rasheed\corref{cor1}}
\ead{zeeshan.rasheed@tuni.fi}

\author[tampere]{Muhammad Waseem}
\ead{muhammad.waseem@tuni.fi}

\author[tampere]{Kai-Kristian Kemell}
\ead{kai-kristian.kemell@tuni.fi}

\author[tampere]{Mika Saari}
\ead{mika.saari@tuni.fi}

\author[tampere]{Pekka Abrahamsson}
\ead{pekka.abrahamsson@tuni.fi}

% ---- Affiliation ----
\affiliation[tampere]{
  organization={Faculty of Information Technology and Communication Sciences, Tampere University},
  city={Tampere},
  country={Finland}
}

% ---- Notes ----
\tnotetext[1]{Submitted to the \textit{Journal of Systems and Software} (JSS).}
\cortext[cor1]{Corresponding author: Zeeshan Rasheed (zeeshan.rasheed@tuni.fi).}

% ---- Abstract & Keywords ----
\begin{abstract}
Large Language Models (LLMs) have enabled multi-agent systems to perform autonomous code generation for complex tasks. Despite the recent growth in research and industrial applications in this area, there is little work on synthesizing evidence from both academic and industrial sources to capture the current state of research on LLM-based multi-agent systems for code generation. To this end, we conducted a Multi-Vocal Literature Review (MLR), combining insights from both academia and industry, including peer-reviewed studies and grey literature. The aim of this study is to systematically synthesize and analyze existing knowledge on LLM-based multi-agent systems for code generation. Specifically, the review examines the motivations for their use, employed benchmarks and models, key challenges, proposed solutions, and potential directions for future research. 

We selected and reviewed 114 studies, and the key findings are: 1) the identified reasons for adopting multi-agent systems for code generation were classified into nine categories; 2) the models and evaluation benchmarks utilized across the studies were systematically analyzed to provide a structured overview of commonly adopted LLM configurations and assessment practices; 3) the reported challenges and corresponding solutions were synthesized into six main categories and 26 subcategories; and 4) future research directions were identified and organized into six main categories and 18 subcategories. The results of this MLR will assist researchers and practitioners in pursuing further studies and supporting the real-world adoption of multi-agent systems in industrial settings.

% We selected and reviewed 114 studies and found that: 1) the identified reasons for adopting multi-agent systems for code generation were classified into nine categories; 2) the models and evaluation benchmarks utilized across the studies were systematically analyzed to provide a structured overview of commonly adopted LLM configurations and assessment practices; 3) the reported challenges and corresponding solutions were synthesized into six main categories and 26 subcategories; and 4) future research directions for multi-agent systems in code generation were identified and organized into six main categories and 18 subcategories. The results of this MLR will assist researchers and practitioners in pursuing further studies and supporting the real-world adoption of multi-agent systems in industrial settings.
\end{abstract}


\maketitle

% ---- Main matter ----
\section{Introduction}
\label{Introduction}

 Large Language Model (LLM)–based agents have been widely adopted in software development processes \cite{belzner2023large}. In recent years, a growing body of research from both academia and industry has examined the capabilities and impact of LLM-based agents across a range of Software Engineering (SE) tasks \cite{liu2024large}. Modern SE workflows increasingly follow an agent-centric design, in which autonomous or semi-autonomous agents play a central role throughout the software development lifecycle \cite{jin2024llms}. This shift toward automated and collaborative software development is exemplified by agent-based frameworks such as MetaGPT \cite{hong2023metagpt}, ChatDev \cite{qian2024chatdev}, AgentCoder \cite{huang2023agentcoder}, MapCoder \cite{DBLP:conf/acl/IslamAP24}, SWE-Agent \cite{yang2024swe}, and CodeSim \cite{DBLP:conf/naacl/IslamAP25}. These systems commonly employ multi-agent architectures, in which each agent serves as a prompt-driven component with well-defined roles, responsibilities, and behavioral constraints. By decomposing complex development tasks into smaller, coordinated sub-tasks, such frameworks enable functional specialization and distribute the cognitive load across agents \cite{he2025llm}.

Despite their potential, the practical adoption of LLM-based multi-agent systems in SE remains in an early stage of development \cite{he2025llm}. Although both academic research and industrial initiatives have demonstrated notable capabilities, fundamental questions remain regarding the reliability, consistency, and generalizability of LLM-based agents across diverse SE contexts \cite{mohammadi2025evaluation}. A range of technical and operational challenges, including hallucinations, inconsistency, high computational costs, and context-length limitations, continue to constrain the effective deployment of LLM-based agents in real-world settings \cite{he2025llm}.

Moreover, the recent growth of white papers, technical blogs, and industry-driven case studies has resulted in a large and loosely connected body of knowledge. This makes it difficult to combine existing findings, compare different approaches, and develop a clear understanding of the current state of LLM-based agent systems in SE \cite{liu2024large}. Furthermore, both researchers and practitioners still face challenges in fully understanding the practical applicability, limitations, and broader implications of employing LLM-based multi-agent systems in software development \cite{jin2024llms}. To catalyze further research and development, there is a need for a systematic synthesis of the existing body of academic and grey literature to provide a structured overview of how these systems are currently used, their limitations, proposed solutions, and future research directions.

\textbf{Motivation}: This study is part of the MAISA project (2025–2027), funded by Business Finland, which aims to investigate the integration of LLM-based agents into SE. The project brings together academia and industry, involving eight leading Finnish companies, each with distinct requirements and goals related to SE and the adoption of LLM-based multi-agent systems. This study addresses industry needs by examining the role of LLM-based agent systems in software development, including tasks such as code generation, and by assessing their reliability. Recent studies indicate that although LLM-based agents for software development are increasingly being explored in industrial settings, organizations continue to face challenges related to reliability, evaluation practices, and integration into existing workflows \cite{wang2025agents}, \cite{he2025llm}, \cite{yang2024swe}, \cite{liu2024large}. These sources also point to a lack of shared knowledge on how agent-based systems are assessed in practice and what types of issues commonly arise during their use. Building on this broader industry context, this study provides an overview of LLM-based multi-agent performance by identifying the key challenges encountered in practice and summarizing the solutions proposed to mitigate these issues. Our study also aims to provide a clear view of LLM-based agent practices, addressing the needs of both researchers and practitioners by bridging academic insights with industrial perspectives.

To address this knowledge gap and meet the practical needs of the MAISA project consortium, in this study we systematically identify and analyze the key objectives, reliability considerations, challenges, proposed solutions, and future research directions related to LLM-based multi-agent system for code generation. Through a Multi-Vocal Literature Review (MLR) that integrates both peer-reviewed and grey literature, we provide a systematic and up-to-date overview of the field and offer actionable insights for researchers, practitioners, and decision-makers.

\textbf{Contributions}: This study makes the following key contributions:

\begin{itemize}

\item An overview of LLM-based multi-agent systems for code generation through systematic analysis of both academic and industrial studies.

\item Identification of the motivations for adopting LLM-based agent system for code generation in both academic and industrial contexts.

\item Extraction of the benchmarks and models reported in both academic and industrial studies and analyzed their frequency of use to identify those most commonly adopted in academia and industry. 

\item Identification and categorization of the challenges and corresponding solutions reported in peer-reviewed literature and grey literature. 

\item Identification of emerging trends and future directions, and recommendations to guide future studies as well as practical adoption in both academic and industrial research on agent-based software development.

\item The replication package of our MLR is publicly available online to support validation and replication of our study \cite{rasheed_2026_18763362}. The dataset includes detailed study demographics, the extracted and analyzed benchmarks and models, as well as the identified and categorized challenges, corresponding solutions, and future research directions for multi-agent systems in software development.

\end{itemize}

\textbf{Structure of the paper}: The structure of the paper is organized as follows. Section \ref{Research Method} presents the research methodology, and the findings of this study are reported in Section \ref{results}. Section \ref{Discussions} discusses the key results and their implications. Section \ref{Threats to Validity} outlines the threats to validity, and Section \ref{background} reviews existing survey studies on LLMs and LLM-based agents in SE. Section \ref{Conclusions} concludes the paper and suggests directions for future work.

% \begin{figure*}[t]
%     \centering
%     \includegraphics[width=1.0\textwidth]{MLR_Eventual_.drawio.pdf}
%     \caption{Visual representation of the research methodology implemented in this study for MLR}
%     \label{fig:placeholder}
% \end{figure*}

% \section{Methodology}
% \label
\section{Research Method}
\label{Research Method}

\begin{figure*}[t]
    \centering
    \includegraphics[width=1.0\textwidth]{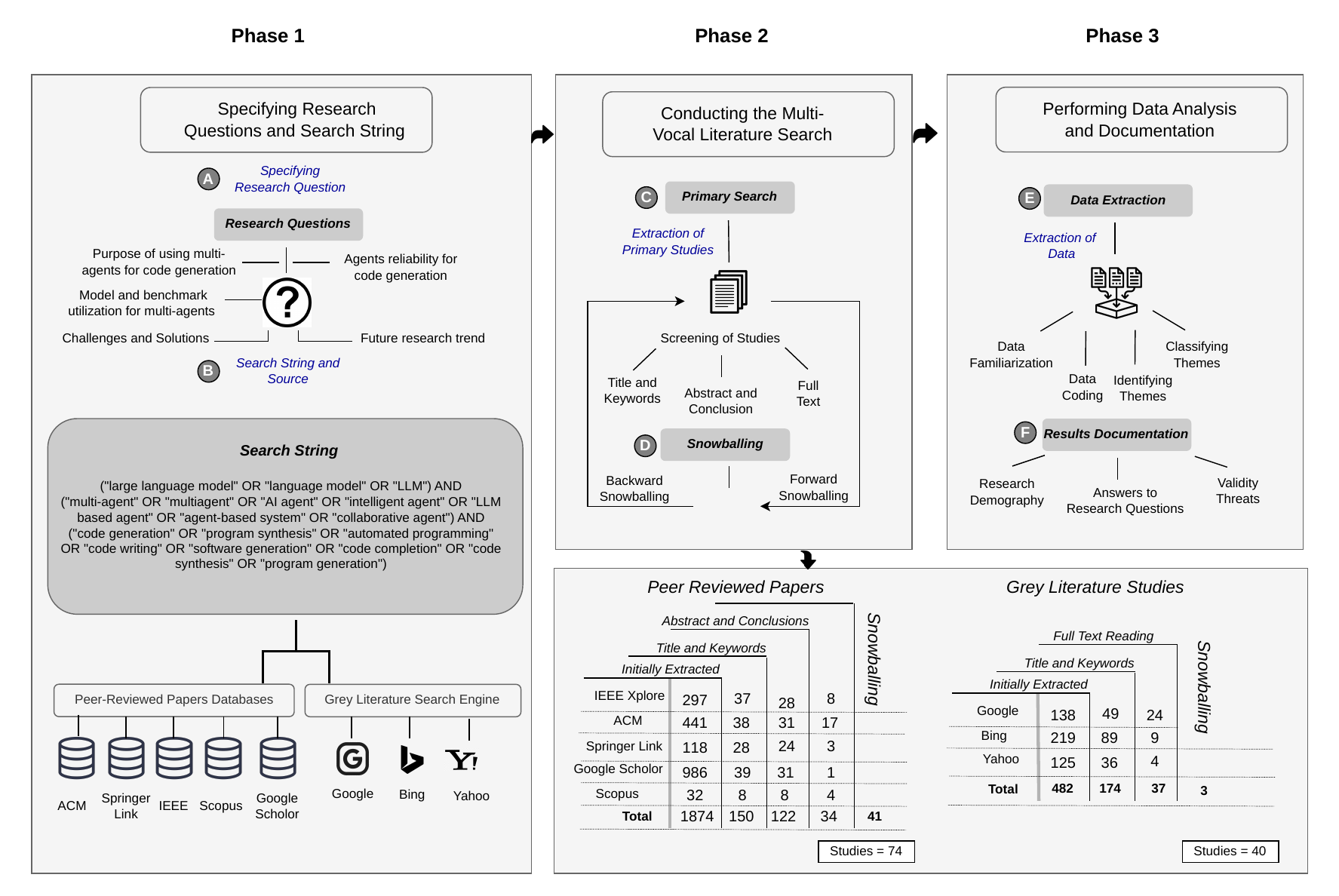}
    \caption{Visual representation of the research methodology implemented in this study for MLR}
    \label{fig:placeholder}
\end{figure*}

We conducted an MLR following the guidelines given in \cite{garousi2019guidelines} and \cite{kitchenham2009systematic}. In an MLR, peer-reviewed academic studies and grey literature are systematically analyzed to provide a broad understanding of a research topic by incorporating perspectives from both academia and industry \cite{garousi2019guidelines}. Our MLR consists of three phases: 1) defining the research questions and search string, 2) conducting the peer-reviewed and grey literature search, and 3) performing data extraction and analysis.
Figure \ref{fig:placeholder} illustrates the MLR process.

\subsection{Objective and Research Questions}

This MLR is designed to examine the role of multi-agent systems in code generation by assessing both the current state of the art and emerging trends. As noted by Garousi \textit{et al}. \cite{garousi2019guidelines}, when a research topic is fast-developing and strongly influenced by industrial practice, grey literature should be taken into consideration. In recent years, the development of agentic frameworks and LLM-based coding systems has been continuously evolving and is strongly driven by industry practice, where many key insights, tools, and empirical observations are reported in grey literature such as technical blogs, open-source repositories, and industry reports. Depending only on peer-reviewed academic studies would not fully represent the current state and practical realities of the field. By integrating evidence from both academic and grey literature, this study aims to offer practical and actionable insights for researchers and practitioners in the SE community. The investigation is structured around seven clearly defined research questions, which are listed in Table \ref{RQs}.

RQ1 maps the evidence base by quantifying the volume and types of publications on multi-agent systems for code generation. Its primary aim is to identify publication trends and the most prominent venues in this research area. RQ2 aims to understand the motivations behind the adoption of multi-agent systems for code generation by researchers and practitioners. RQ3 examines the evaluation benchmarks and metrics used to assess agent performance.
RQ4 focuses on identifying the language models employed in academia and industry for these agent systems. Since language models serve as the core reasoning component of such systems, it is important to determine which models are most frequently adopted. Finally, RQ5, RQ6, and RQ7 collectively address the research gaps: RQ5 analyzes the challenges encountered, RQ6 investigates the proposed solutions to mitigate these challenges and enhance performance, and RQ7 outlines future research directions and open issues identified by the community.
{\renewcommand{\arraystretch}{1}
\begin{table*}[t]
\centering
\scriptsize
\caption{Research questions and rationales for the MLR on LLM-based multi-agent systems for code generation}
\label{RQs}

\resizebox{\textwidth}{!}{%
\begin{tabular}{|l|l|l|}
\hline
\textbf{RQ} & \textbf{Research Question} & \textbf{Rationale} \\ \hline

\multicolumn{3}{|c|}{\textbf{Demographic Detail}} \\ \hline
RQ1 &
\begin{tabular}[c]{@{}l@{}}
What is the frequency and type of published research on LLM-based \\
multi-agent systems for code generation?
\end{tabular} &
\begin{tabular}[c]{@{}l@{}}
This RQ aims to map the evidence base by collecting data on the volume and type \\
of literature concerning multi-agent systems for code generation. The findings derived \\
from this RQ will establish publication trends and identify the most frequent \\
publishing venues for multi-agent system research in this domain.
\end{tabular} \\ \hline

\multicolumn{3}{|c|}{\textbf{Reasons for using Agents}} \\ \hline
RQ2 &
\begin{tabular}[c]{@{}l@{}}
What are the main purposes for using multi-agent LLM-based systems \\
in code generation?
\end{tabular} &
\begin{tabular}[c]{@{}l@{}}
This question aims to identify why LLM-based multi-agent systems are adopted for \\
code generation, particularly in addressing complexity, verification, and quality \\
issues. It helps clarify the practical motivations behind using multiple agents instead \\
of single-agent approaches.
\end{tabular} \\ \hline

\multicolumn{3}{|c|}{\textbf{Model and Benchmark Utilization}} \\ \hline
RQ3 &
\begin{tabular}[c]{@{}l@{}}
What evaluation metrics or benchmarks are used to assess these LLM-agent \\
systems in code generation?
\end{tabular} &
\begin{tabular}[c]{@{}l@{}}
The objective of this question is to identify the evaluation benchmarks and metrics \\
used to assess LLM-based multi-agent systems in code generation, clarifying how multi-agent \\
performance, quality, and reliability are measured across studies.
\end{tabular} \\ \hline

RQ4 &
\begin{tabular}[c]{@{}l@{}}
What LLM models are most commonly employed in code generation tasks?
\end{tabular} &
\begin{tabular}[c]{@{}l@{}}
This question examines which LLM models are most frequently used in \\
code generation to reveal common choices and trends. It helps identify \\
standard practices and informs comparisons across existing multi-agent systems.
\end{tabular} \\ \hline

\multicolumn{3}{|c|}{\textbf{Challenges, Solutions, and Future Research Trends}} \\ \hline
RQ5 &
\begin{tabular}[c]{@{}l@{}}
What are the main challenges reported in applying LLM-based \\
multi-agent systems for code generation?
\end{tabular} &
\begin{tabular}[c]{@{}l@{}}
The objective of this question is to identify the main challenges reported \\
in the application of multi-agent systems for code generation.
\end{tabular} \\ \hline

RQ6 &
\begin{tabular}[c]{@{}l@{}}
What solutions are proposed to mitigate these challenges and improve the \\
performance of LLM-based multi-agent systems for code generation?
\end{tabular} &
\begin{tabular}[c]{@{}l@{}}
The objective of this question is to examine the solutions proposed to address \\
identified challenges and to improve the performance of LLM-based multi-agent \\
systems for code generation.
\end{tabular} \\ \hline

RQ7 &
\begin{tabular}[c]{@{}l@{}}
What future research directions and open challenges are identified in studies \\
on LLM-based multi-agent systems for code generation?
\end{tabular} &
\begin{tabular}[c]{@{}l@{}}
The objective of this question is to identify future research directions and open \\
challenges highlighted in studies on LLM-based multi-agent systems for code \\
generation, emphasizing gaps, emerging trends, and areas requiring further \\
investigation.
\end{tabular} \\ \hline

\end{tabular}%
}
\end{table*}}

\subsection{Search Strategy}
\label{Search Strategy}

In this section, we describe the process used to collect both peer-reviewed studies and grey literature sources for inclusion in our review. 

\subsubsection{Search String Development}

First, we considered using the PICO (Population, Intervention, Comparison, Outcome) framework \cite{schardt2007utilization} to guide the development of the search strategy. However, applying PICO resulted in an extensive combination of keywords and Boolean operators, which exceeded the query length limitations imposed by several digital libraries. Finally, we formulated the search string using our domain knowledge and a series of iterative trial searches conducted with the co-authors. This approach allowed us to refine the search strategy to align with the aims of our study while maintaining broad coverage within practical constraints.

In the development of our search string, we combined domain-specific terminology to capture studies related to LLMs and their application in multi-agent and code-generation contexts. Specifically, we used combinations of the phrases ``large language model'', ``language model'', and ``LLM'' to represent general references to foundation models. To capture literature focused on multi-agent collaboration, we included variations such as ``multi-agent'', ``multiagent'', ``AI agent'', ``intelligent agent'', ``LLM-based agent'', ``agent-based system'', and ``collaborative agent''. Finally, to target research addressing SE applications, the search string incorporated phrases such as ``code generation'', ``program synthesis'', ``automated programming'', ``code writing'', ``software generation'', ``code completion'', and ``program generation''. Logical operators (AND/OR) were applied to ensure that all retrieved studies discussed the intersection of these core themes.

{\renewcommand{\arraystretch}{1}
\begin{table}[t]
\centering
\scriptsize
\caption{Sources for collecting peer-reviewed and grey literature}
\label{tab:sources}

\begin{tabular}{|p{0.42\columnwidth}|p{0.48\columnwidth}|}
\hline
\multicolumn{2}{|c|}{\textbf{Databases (Peer-reviewed / White Literature)}} \\ \hline

\textbf{Database} & \textbf{Targeted Search Area} \\ \hline
ACM Digital Library & Paper title, keywords, abstract \\ \hline
IEEE Xplore         & Paper title, keywords, abstract \\ \hline
Scopus              & Paper title, keywords, abstract \\ \hline
SpringerLink        & Paper title, abstract \\ \hline
Google Scholar      & Paper title, abstract \\ \hline

\multicolumn{2}{|c|}{\textbf{Grey Literature Sources}} \\ \hline

\textbf{Source} & \textbf{Targeted Search Area} \\ \hline
Google & Web search (titles/snippets); domain filters when relevant \\ \hline
Bing   & Web search (titles/snippets); domain filters when relevant \\ \hline
Yahoo  & Web search (titles/snippets); domain filters when relevant \\ \hline

\end{tabular}
\end{table}}

% \begin{table*}[t]
% \centering
% \caption{Search String and Sources Consulted}
% \label{tab:sources}
% \begin{tabular}{p{0.35\textwidth} p{0.6\textwidth}}
% \hline
% \multicolumn{2}{c}{\textbf{Search String}}\\
% \multicolumn{2}{p{0.95\textwidth}}{\centering (\texttt{("large language model" OR "language model" OR "LLM") AND ("multi-agent" OR "multiagent" OR "AI agent" OR "intelligent agent" OR "LLM-based agent" OR "agent-based system" OR "collaborative agent") \\ AND ("code generation" OR "program synthesis" OR "automated programming" OR "code writing" OR "software generation" OR "code completion" OR "code synthesis" OR "program generation")})}\\
% \hline
% \multicolumn{2}{c}{\textbf{Databases (peer-reviewed / white literature)}}\\
% \textbf{Database} & \textbf{Targeted search area} \\
% \hline
% ACM Digital Library & Paper title, keywords, abstract \\
% IEEE Xplore          & Paper title, keywords, abstract \\
% Scopus               & Paper title, keywords, abstract \\
% SpringerLink         & Paper title, abstract \\
% Google Scholar  & Paper title, abstract \\
% \hline
% \multicolumn{2}{c}{\textbf{Grey literature}}\\
% \textbf{Source} & \textbf{Targeted search area} \\
% \hline
% Google & Web search (titles/snippets); use site/domain filters where relevant \\
% Bing   & Web search (titles/snippets); use site/domain filters where relevant \\
% Yahoo  & Web search (titles/snippets); use site/domain filters where relevant \\
% \hline
% \end{tabular}
% \end{table*}

\subsubsection{Bibliographic Sources}
To retrieve white papers, we selected relevant bibliographic sources in accordance with the guidelines proposed by Kitchenham and Charters \cite{brereton2007lessons}. For the white literature, we searched five major sources: ACM Digital Library, IEEE Xplore, Scopus, SpringerLink, and Google Scholar. For the grey literature, we used three widely used search engines: Google, Bing, and Yahoo.

\subsubsection{Inclusion and Exclusion Criteria}

To maintain the quality and relevance of the reviewed studies, a set of inclusion and exclusion criteria was defined according to the objectives of this study (see Table~\ref{Tab:Inc and Exc}). Studies were included if they (1) investigated LLM-based multi-agent systems for code generation, (2) reported an evaluation of agent-based systems for code generation using benchmarks such as HumanEval, SWE-bench, APPS, MBPP, or other measurable real-world programming tasks, and (3) were published in English as peer-reviewed papers or credible grey literature within the defined temporal scope. In contrast, studies were excluded if they (1) were not available in English or lacked full-text accessibility through institutional access, (2) fell outside the scope of LLM-based agentic systems for code generation, or (3) consisted of opinion pieces, news articles, or short abstracts that did not provide sufficient methodological detail or evaluative evidence.

% \begin{table*}[!htbp]
% \centering
% \caption{Inclusion and exclusion criteria defined for this MLR}
% \label{Tab:Inc and Exc}

% \begin{tabularx}{0.95\textwidth}{l X X}
% \toprule
% \textbf{Selection Criteria} & \textbf{Inclusion Criteria} & \textbf{Exclusion Criteria} \\
% \midrule

% \textit{Language} &
% English &
% Non-English \\
% \midrule

% \textit{Study type} &
% Primary studies from peer-reviewed sources (e.g., journal articles, conference papers, workshop and symposium papers) as well as relevant grey literature (e.g., technical reports, practitioner blogs, industry documents, videos). &
% Duplicate studies or sources lacking accessible full text (e.g., broken links, missing documents). \\
% \midrule

% \textit{Study focus} &
% Studies that explicitly focus on LLM-based agents for code generation. This includes research on agent-based programming workflows, the development or application of LLM-driven agents for automating code generation, and evaluations of agent performance in code generation tasks. &
% Studies that discuss LLM-based agents without involving code generation tasks (e.g., reasoning, planning, dialogue), studies using LLMs for code generation without an agent-based component, and studies on agents performing tasks unrelated to code generation. \\
% \midrule

% \textit{Study duration} &
% Studies published between January 2022 and June 2025. &
% Studies published before January 2022 or after June 2025. \\
% \bottomrule
% \end{tabularx}

% \end{table*}
{\renewcommand{\arraystretch}{1}
\begin{table*}[t]
\centering
\scriptsize
\caption{Inclusion and exclusion criteria defined for this MLR}
\label{Tab:Inc and Exc}

\begin{tabularx}{1.0\textwidth}{|l|X|X|}
\hline
\textbf{Selection Criteria} & \textbf{Inclusion Criteria} & \textbf{Exclusion Criteria} \\ \hline

\textit{Language} &
English &
Non-English \\ \hline

\textit{Study type} &
Primary studies from peer-reviewed sources (e.g., journal articles, conference papers, workshop and symposium papers) as well as relevant grey literature (e.g., technical reports, practitioner blogs, industry documents, videos). &
Duplicate studies or sources lacking accessible full text (e.g., broken links, missing documents). \\ \hline

\textit{Study focus} &
Studies that explicitly focus on LLM-based agents for code generation. This includes research on agent-based programming workflows, the development or application of LLM-driven agents for automating code generation, and evaluations of agent performance in code generation tasks. &
Studies that discuss LLM-based agents without involving code generation tasks (e.g., reasoning, planning, dialogue), studies using LLMs for code generation without an agent-based component, and studies on agents performing tasks unrelated to code generation. \\ \hline

\textit{Study duration} &
Studies published between January 2022 and June 2025. &
Studies published before January 2022 or after June 2025. \\ \hline

\end{tabularx}
\end{table*}}

{\renewcommand{\arraystretch}{1}
\begin{table*}[!t]
\centering
\scriptsize
\caption{Quality assessment criteria}
\label{Grey Literature QA}
% \resizebox{\textwidth}{!}{%
% \begin{tabular}{|l|l|l|}
\begin{tabularx}{\linewidth}{|p{0.18\linewidth}|X|X|}

\hline
\multicolumn{3}{|c|}{\textbf{Quality Assessment Criteria for Peer-Reviewed Studies}} \\ \hline
\textbf{QA} & \multicolumn{2}{l|}{\textbf{Descriptions}} \\ \hline
QA1 & \multicolumn{2}{l|}{Is the study focused on the use of LLM-based agents for code generation tasks?} \\ \hline
QA2 & \multicolumn{2}{l|}{Is there a clear statement of the aims of the research?} \\ \hline
QA3 & \multicolumn{2}{l|}{Is the study design suitable for achieving the stated research objectives?} \\ \hline
QA4 & \multicolumn{2}{l|}{Are empirical or experimental data available to support the research?} \\ \hline
QA5 & \multicolumn{2}{l|}{Was the data collection process aligned with the research objectives?} \\ \hline
QA6 & \multicolumn{2}{l|}{Was the data analysis conducted in a systematic and robust manner?} \\ \hline
QA7 & \multicolumn{2}{l|}{Has the relationship between researcher and participants been appropriately considered?} \\ \hline
QA8 & \multicolumn{2}{l|}{Is there a clear statement of findings?} \\ \hline
QA9 & \multicolumn{2}{l|}{Is the study valuable for research or practice?} \\ \hline
\multicolumn{3}{|c|}{\textbf{Quality Assessment for Grey Literature}} \\ \hline
\textbf{Criteria} & \textbf{Questions} & \textbf{Possible Answers} \\ \hline
 & Is the publishing organization reputable? & \begin{tabular}[c]{@{}l@{}}1: Well-known organization\\ 0.5: Existing but not well known \\ 0: Low-reputation organization\end{tabular} \\ \cline{2-3}
 & Is the author affiliated with a reputable organization? & \begin{tabular}[c]{@{}l@{}}1: Yes\\ 0: No\end{tabular} \\ \cline{2-3}
\multirow{-3}{*}{\begin{tabular}[c]{@{}l@{}}Author and Source \\ Credibility\end{tabular}} & Does the author have sufficient experience in the field? & \begin{tabular}[c]{@{}l@{}}1: More than three publications\\ 0.5: 1–2 publications\\ 0: No prior publications\end{tabular} \\ \hline
 & Does the source have a clearly stated aim? & \begin{tabular}[c]{@{}l@{}}1: Yes\\ 0: No\end{tabular} \\ \cline{2-3}
 & Is the source supported by authoritative references? & \begin{tabular}[c]{@{}l@{}}1: Reputable sources\\ 0.5: Less reputable sources\\ 0: No references\end{tabular} \\ \cline{2-3}
 & Does the work cover a specific research question? & \begin{tabular}[c]{@{}l@{}}1: Yes\\ 0.5: Partially\\ 0: No\end{tabular} \\ \cline{2-3}
\multirow{-4}{*}{Research Objectivity} & Are the conclusions free from bias? & \begin{tabular}[c]{@{}l@{}}1: No vested interest\\ 0.5: Minor interest\\ 0: Strong interest\end{tabular} \\ \hline
 & Are the conclusions supported by data? & \begin{tabular}[c]{@{}l@{}}1: Yes\\ 0.5: Partially\\ 0: No\end{tabular} \\ \cline{2-3}
 & Does the item have a clearly stated publication date? & \begin{tabular}[c]{@{}l@{}}1: Yes\\ 0: No\end{tabular} \\ \cline{2-3}
\multirow{-3}{*}{Evidence and Date} & Are key grey or formal sources discussed? & \begin{tabular}[c]{@{}l@{}}1: Yes\\ 0: No\end{tabular} \\ \hline
Context and Novelty & Does the source contribute novel insights? & \begin{tabular}[c]{@{}l@{}}1: Yes\\ 0.5: Partially\\ 0: No\end{tabular} \\ \hline
Outlet Type & Level of outlet control & \begin{tabular}[c]{@{}l@{}}1: High control (books, reports, etc.)\\ 0.5: Moderate control (articles, Q/A sites)\\ 0: Low control (blogs, emails, tweets)\end{tabular} \\ \hline
\end{tabularx}
%}
\end{table*}}

\subsubsection{Search and Selection Process for Peer-Reviewed and Grey Literature}

We conducted the search and selection process for peer-reviewed papers in June 2025 and considered all publications released between 2022 and 2025. The application of our search terms resulted in 1,874 unique papers, as shown in Figure \ref{fig:placeholder}. For grey literature, the search was conducted in September 2025 and included all publications available up to that date. Applying the search terms resulted in 482 unique grey literature contributions, as reported in Figure \ref{fig:placeholder}.

\paragraph{Primary Search:}

\begin{itemize}
    \item Step 1: Extraction of Studies:
    For collecting \textbf{peer-reviewed studies}, we performed custom search queries across the selected databases (see Table \ref{tab:sources}) to retrieve study titles, authors, publication years, venues, publication types, and abstracts. This initial retrieval process resulted in 1,874 studies collected from five databases.
    \textbf{For grey literature}, we used the same search terms as those adopted for peer-reviewed studies. We performed custom search queries across three search engines, Google, Bing, and Yahoo, resulting in 482 studies retrieved from these sources. 
    %Because this step was entirely automated and involved no subjective decision-making, Cohen’s Kappa \cite{cohen1960coefficient} was not applicable for assessing inter-rater reliability.
    
    \item Step 2: Title and Keyword Screening:
    \textbf{For peer-reviewed studies}, we manually reviewed the titles and keywords of the papers initially collected using the predefined inclusion and exclusion criteria shown in Table~\ref{Tab:Inc and Exc}. We included only studies that examined LLM-based agents for code generation or software development, while papers focusing solely on multi-agent or agent-based systems without a connection to code generation were excluded. Duplicate entries collected from multiple databases (e.g., IEEE Xplore, ACM, and Scopus) were identified and removed by sorting and comparing records, resulting in the elimination of 478 duplicates from the initial set of 1,874 papers. After applying the inclusion and exclusion criteria to titles and keywords, 150 papers were retained for the next stage of the review.
    \textbf{For grey literature}, duplicate studies retrieved from Google, Bing, and Yahoo were first identified and removed by sorting and comparing records, resulting in the elimination of 84 studies from the initial set of 482 studies. Next, the titles and abstracts of the remaining 398 studies were manually screened using the inclusion and exclusion criteria presented in Table~\ref{Tab:Inc and Exc}. This screening process covered a variety of sources, including blogs, technical reports, arXiv papers, articles, and videos. As a result, a total of 178 studies were selected for full-text screening.

    \item Step 3: Abstract-Based Screening:
    \textbf{For peer-reviewed studies}, we reviewed the abstracts of the collected papers to assess their alignment with our research topic. The first author independently examined the abstract of each paper and assigned a status of ``Yes'' or ``No'' based on its relevance to the research objectives.  Following this process, 122 papers were retained after applying the abstract-level inclusion and exclusion criteria. \textbf{For grey literature}, most sources do not provide an abstract; therefore, we skipped the abstract screening stage and directly reviewed the full text.

    \item Step 4: Full-Text Screening:
    
    \textbf{For peer-reviewed studies}, we reviewed the full texts of the 122 papers shortlisted during the abstract screening stage. After applying the inclusion and exclusion criteria to the full-text assessments, we selected 33 papers for final data extraction. \textbf{For grey literature}, we conducted a full-text review of the 174 articles selected during the title screening stage, applying the criteria defined in Table~\ref{Tab:Inc and Exc}. Based on this assessment, we identified 39 relevant studies: 18 blog posts, nine arXiv papers, five technical reports, two YouTube videos, two news articles, two media posts, and one handbook.

% To ensure consistency in our review process, we randomly selected a
% subset of 35 studies (10\% of the total 346 studies) for
% a Cohen’s Kappa analysis to measure the agreement
% between two researchers (R1 and R2). The results,
% recorded in Table 4.c, indicated a Cohen’s Kappa
% value of k = 0.578, which represents moderate agreement between the two researchers [23]. Specifically,
% R1 and R2 agreed on 30 studies as relevant (“YES”)
% and 2 studies as irrelevant (“NO”). Disagreements
% occurred in 3 studies (2 marked as “YES” by one
% researcher and “NO” by the other, and 1 marked as
% “NO” by both but flagged as uncertain). All disagreements were systematically resolved through discussions among the authors, leading to a final consensus.
% This subset-based approach, coupled with systematic
% resolution of disagreements, provides confidence in
% the reliability of the full-text screening process [24].
% At the end of this step, we obtained a total of 110 studies, which were deemed relevant for further analysis.
\end{itemize}

\paragraph{Snowballing:}

In Phase 2, we applied a snowballing technique, following the general principles outlined by Wohlin \cite{wohlin2014guidelines}, to expand our set of primary studies. This involved reviewing the reference lists of 33 selected \textbf{peer-reviewed studies} (backward snowballing) and identifying new studies that cited these papers (forward snowballing). The snowballing process was conducted in August 2025. Through this process, we identified an additional 34 studies through backward snowballing and seven studies through forward snowballing. In total, 74 studies were included for data extraction and analysis (41 from snowballing and 33 from database searches).
\textbf{For grey literature}, we also applied both backward and forward snowballing, which was conducted in November 2025. In total, three articles were identified and included in the final set of publications: two obtained through backward snowballing and one through forward snowballing.

{\renewcommand{\arraystretch}{1}
\begin{table*}[t]
\centering
\scriptsize
\caption{Data items to be extracted in this MLR}
\label{Tab Dataitem}

\begin{tabular}{|l|l|p{12cm}|l|}
\hline
\textbf{Code} & \textbf{Data Item} & \textbf{Description} & \textbf{RQ} \\ \hline

D1 & Index & ID of the study & \multirow{6}{*}{RQ1} \\ \cline{1-3}
D2 & Publication Year & Publication year of the study & \\ \cline{1-3}
D3 & Publisher & Publisher of the study & \\ \cline{1-3}
D4 & Venue & Name of the publishing venue & \\ \cline{1-3}
D5 & Publication Type & Journal, conference, workshop, and book chapter & \\ \cline{1-3}
D6 & Authors’ Affiliation & Affiliation of the authors & \\ \hline

D7  & Purpose of using Agents & Purpose of using agents for code generation in the study & {RQ2} \\ \hline
D8  & Benchmark and Metrics   & Benchmarks and metrics used for testing agents performance in the study & {RQ3} \\ \hline
D9  & Large Language Models   & Large language model used for agents & {RQ4} \\ \hline
D10 & Challenges              & Agents' challenges regarding code generation in the study & {RQ5} \\ \hline
D11 & Solutions               & Solutions for the agents' challenges in the study & {RQ6} \\ \hline
D12 & Future Research Trend   & Potential future research trend for agents in the study & {RQ7} \\ \hline

\end{tabular}
\end{table*}}

\paragraph{Quality Assessment of Selected Studies:}

\textbf{For peer-reviewed studies}, we evaluated the quality of the selected studies by following the guidelines proposed by Dybå \textit{et al}. \cite{dybaa2008empirical}. First, we developed a checklist (Table~\ref{Grey Literature QA}) consisting of specific evaluation questions to ensure the relevance of the studies to our research topic. Each response was rated using a five-point Likert scale, ranging from 0 (poor) to 4 (excellent). The assessment considered each study’s focus on LLM-based agents for code generation, the clarity and structure of its research design, the availability and strength of empirical evidence (e.g., experiments or case studies), the quality of reporting (including discussion of limitations), publication status in peer-reviewed venues, and the extent to which the findings could be generalized to multi-agent systems for code generation. The quality assessment was performed after the study selection process to ensure a consistent and thorough evaluation of all 74 included studies. 
Across the included peer-reviewed studies, quality scores ranged from 0 (poor) to 0.5 (average) and 1 (excellent). Most included studies achieved moderate to high quality scores, suggesting that the synthesized findings are based on methodologically sound evidence.

Unlike peer-reviewed publications, \textbf{grey literature} is not formally peer-reviewed, which results in less quality control.
To assess the credibility and quality of grey literature sources and to determine their inclusion, we followed the guidelines proposed by Garousi \textit{et al}. \cite{garousi2019guidelines}. As shown in Table~\ref{Grey Literature QA}, the evaluation considers factors such as author and source credibility, research objectives, data credibility, novelty, impact, and the level of outlet control. Each source was initially evaluated by the first author using the defined criteria, applying either a binary scale or a three-point Likert scale as appropriate. The second and third authors then conducted independent reviews of these evaluations. Any disagreements were resolved through discussion. The final decision was based on the average score, and grey literature sources with a score below 0.5 on a scale from 0 to 1 were excluded.
We finally applied the quality assessment to the 39 selected studies, excluding two studies and retaining 37 studies for the subsequent data extraction phase.

\subsection{Data Extraction}

Initially, we collected 2,356 unique studies (1,874 white literature and 482 grey literature sources); the selection process resulted in 114 final studies (74 white literature and 40 grey literature sources), as presented in Figure \ref{fig:placeholder}.

The data extraction form was developed based on a predefined set of data items (see Table~\ref{Tab Dataitem}), each aligned with the RQs presented in Table~\ref{RQs}. To minimize potential bias and ensure a shared understanding, the authors jointly discussed and cross-verified the extracted results. As shown in Table~\ref{Tab Dataitem}, data items D1–D6 capture the general characteristics of the primary studies, whereas data items D7–D12 are directly mapped to RQ2–RQ7. A description of each data item is provided in Table~\ref{Tab Dataitem}, and all extracted data were systematically organized and analyzed using Excel spreadsheets.

\subsection{Data Analysis}

In this study, we employed a mixed-methods approach to analyze the extracted data, using descriptive analysis to summarize quantitative results and thematic analysis to synthesize qualitative findings. Quantitative data corresponding to data items D1, D2, D3, D4, D5, D6, D8, and D9 were analyzed using descriptive statistics \cite{wohlin2006empirical} to provide an overview of the study characteristics. The remaining data consisted of qualitative information, such as reasons for adopting agents, identified challenges and their proposed solutions, and directions for future research. These qualitative data were examined using a thematic analysis-inspired coding approach to organize the data into categories and subcategories relevant to the research questions, informed by the guidelines presented in \cite{terry2017thematic}. The analysis was conducted through the following steps:

\begin{itemize} 

    \item \textbf{Data Overview:} All included studies were reviewed multiple times to develop a thorough understanding of their content. During this process, we systematically identified and documented key information related to the purpose of using agents (D7), reported challenges and proposed solutions (D10–D11), and future research trends (D12).

    \item \textbf{Initial Code Generation:} The next step involved developing an initial coding scheme based on the information extracted for the above mentioned data items.

    \item \textbf{Emerging Themes:} In this phase, a two-stage analysis was conducted. Initially, the identified codes were reviewed to explore their nature and interrelationships. Then, related codes were grouped into subcategories and broader thematic categories representing higher-level concepts across the studies.

    \item \textbf{Analytical Review:} After the analysis of the extracted data, all authors collaboratively engaged in the review and refinement of the coded data, including the identified types, subcategories, and categories. Through iterative discussions, some themes were revised, merged, or removed to improve conceptual clarity and ensure consistency across the analysis.

    \item \textbf{Theme Definition:} Finally, clear definitions were developed for each finalized theme. The terminology and naming conventions were carefully refined to improve clarity, readability, and alignment with the objectives of the study.
\end{itemize}

\section{Results}
\label{results}
In this section, we present the results of the MLR, derived from the analysis of selected literature. Section~\ref{Demographics} reports the demographic information of both peer-reviewed and grey literature, including publication year, publisher or organization, and authors’ affiliations. Sections~\ref{Reasons} and~\ref{Benchmakr and Model Distribution} examine the reasons for adopting LLM-based agents for code generation and analyze the benchmarks and language models most commonly used in agent-based systems.

%Section~\ref{Code Reliability} reports findings related to the reliability of code generated by agents, while 
Section~\ref{chalengess and Solutions} summarizes the challenges faced by LLM-based agent systems and the corresponding solutions proposed in the literature for code generation tasks. Finally, Section~\ref{Future Work} outlines the directions for future research identified across the reviewed studies.

\subsection{Demographics of Peer-Reviewed and Grey Literature (RQ1)}
\label{Demographics}

\begin{figure*}[!t]
    \centering
    \includegraphics[width=\textwidth]{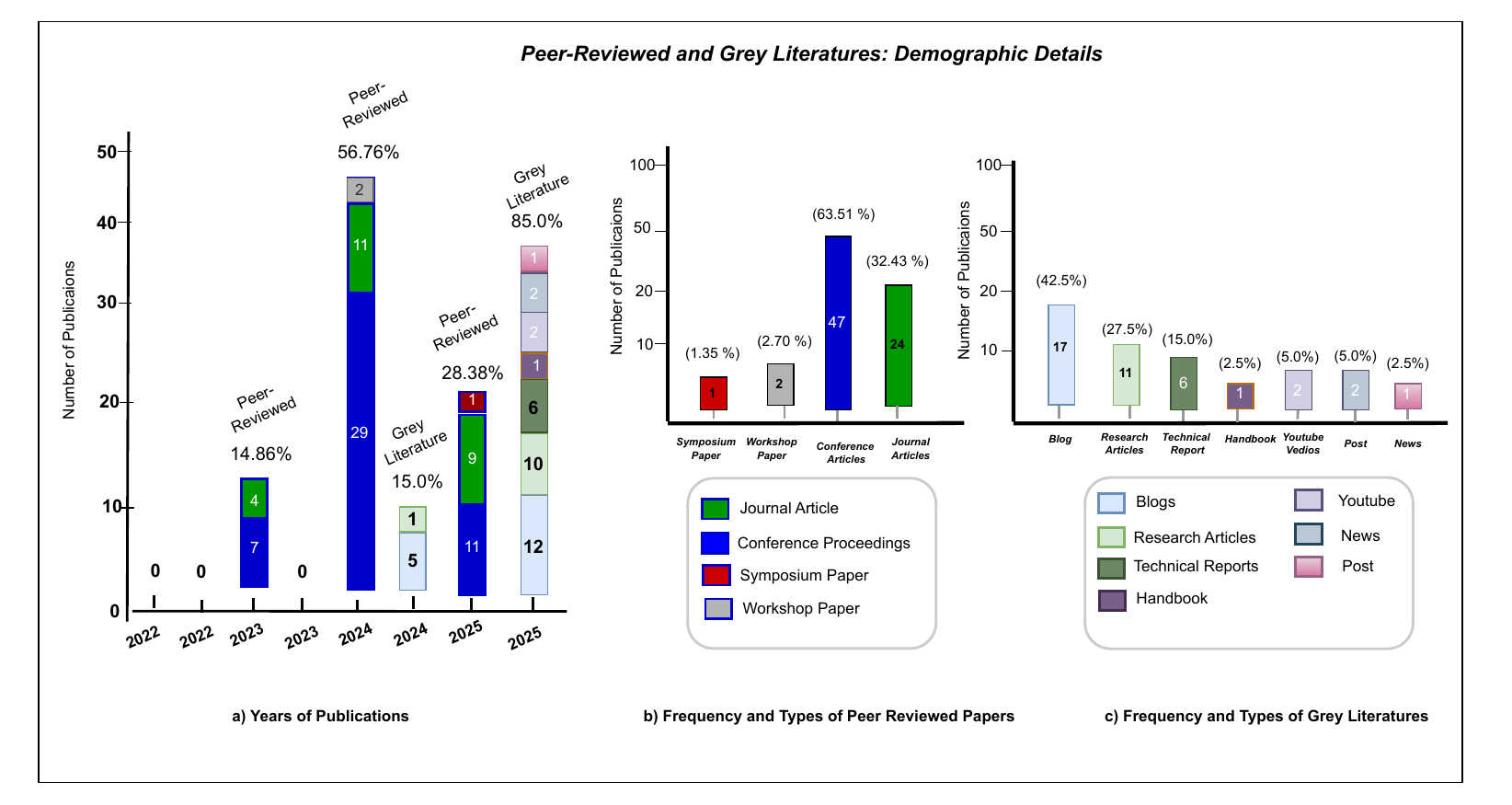}
    \caption{Demographic distribution of peer-reviewed and grey literature studies}
    \label{fig:demography_white}
\end{figure*}

%\subsubsection{Peer-Reviewed Papers Demographics}
\subsubsection{Yearly Distribution of Papers} Figure~\ref{fig:demography_white}a shows the annual distribution of peer-reviewed and grey literature studies. \textbf{Peer-reviewed studies} were collected from five databases between January 2022 and June 2025. No studies on LLM-based agents for code generation were identified in 2022. Although the first agent-based systems were introduced in 2022, they did not target code generation tasks. In 2023, a total of 11 studies were published, comprising seven conference papers and four journal articles.
As shown in Figure~\ref{fig:demography_white}a, there was a notable increase in research activity in 2024, with 42 studies published. Of these, 29 were conference papers, 11 were journal articles, and two were workshop papers. By June 2025, 21 studies had been published, including 11 conference papers, nine journal articles, and one symposium paper.

\textbf{Grey literature} was collected from three sources between January 2022 and September 2025. We did not find any paper between 2022 and 2023 that specifically focused on LLM-based agent systems for code generation. While broader concepts of autonomous agents and LLMs were emerging during this period, their specialized application to automated SE and code generation tasks had not yet attracted attention within the scope of our review.
As shown in Figure~\ref{fig:demography_white}a, we found six papers for 2024, which shows the start of a growing research focus on LLM-based agents for code generation. These papers consist of one research article and five blog posts. In 2025, we witnessed a peak in research activity, with 34 papers. This period shows a more diverse range of publication types, including 11 research articles, six technical reports, twelve blog posts, two videos, two posts, one technical handbook, and one newsroom article.

\begin{table}[t]
\centering
\caption{Distribution of peer-reviewed and grey literature}
\label{tab:literature_distribution}
%\begin{tabular}{lp{4.8cm}c}
%\begin{tabular}{|l p{4.3cm} c|}
\begin{tabular}{|l|p{4.3cm}|c|}
\hline
\textbf{Category} & \textbf{Paper IDs} & \textbf{\%} \\
\hline
\multicolumn{3}{|c|}{\textbf{Peer-Reviewed Studies}} \\
\hline
Peer-reviewed &
\ref{P1}, \ref{P2}, \ref{P3}, \ref{P4}, \ref{P5}, \ref{P6}, \ref{P7}, \ref{P8}, \ref{P9}, \ref{P10}, \ref{P11}, \ref{P12}, \ref{P13}, \ref{P14}, \ref{P15}, \ref{P16}, \ref{P17}, \ref{P18}, \ref{P19}, \ref{P20}, \ref{P21}, \ref{P22}, \ref{P23}, \ref{P24}, \ref{P25}, \ref{P26}, \ref{P27}, \ref{P28}, \ref{P29}, \ref{P30}, \ref{P31}, \ref{P32}, \ref{P33}, \ref{P34}, \ref{P35}, \ref{P36},
\ref{P37}, \ref{P38}, \ref{P39}, \ref{P40}, \ref{P41}, \ref{P42}, \ref{P43}, \ref{P44}, \ref{P45}, \ref{P46}, \ref{P47}, \ref{P48}, \ref{P49}, \ref{P50}, \ref{P51}, \ref{P52}, 
\ref{P53}, \ref{P54}, \ref{P55}, \ref{P56},
\ref{P57}, \ref{P58}, \ref{P59}, \ref{P60}, \ref{P61}, \ref{P62}, \ref{P63}, \ref{P64}, \ref{P65}, \ref{P66}, \ref{P67}, \ref{P68}, \ref{P69}, \ref{P70}, \ref{P71}, \ref{P72}, \ref{P73}, \ref{P74},

& 64.91 \\
\hline
\multicolumn{3}{|c|}{\textbf{Grey Literature}} \\
\hline
Grey Literature &
\ref{G1}, \ref{G2}, \ref{G3}, \ref{G4},  \ref{G5}, \ref{G6},  \ref{G7}, \ref{G8},  \ref{G9}, \ref{G10},  \ref{G11}, \ref{G12},  \ref{G13}, \ref{G14},  \ref{G15}, \ref{G16}, \ref{G17}, \ref{G18},  \ref{G19}, \ref{G20},  \ref{G21}, \ref{G22},  \ref{G23}, \ref{G24},  \ref{G25}, \ref{G26},  \ref{G27}, \ref{G28}, \ref{G29}, \ref{G30},  \ref{G31}, \ref{G32},  \ref{G33}, \ref{G34},  \ref{G35}, \ref{G36},  \ref{G37}, \ref{G38},  \ref{G39}, \ref{G40},

& 35.09 \\
\hline
\end{tabular}
\end{table}

\subsubsection{Study Types}
\textbf{For peer-reviewed studies}, Figure~\ref{fig:demography_white}b presents the distribution of the selected studies by publication type. Conference papers form the largest share of the publications, with a total of 47 studies. Journal articles follow with 24 publications, indicating their role as venues for more detailed and carefully reviewed research. This distribution suggests that conferences are more commonly used to report recent developments, due to faster review processes and quicker sharing of results. Workshops and symposiums account for smaller numbers, with two and one publications, respectively, suggesting that these venues are used less frequently, possibly because they focus on more specialized topics.

%\paragraph{Source Types}
\textbf{For grey literature}, blogs account for 17 of the 40 selected studies, highlighting the active role of industry practitioners and developers in documenting agent-based code generation techniques. Additionally, 12 studies are research articles hosted on arXiv. As shown in Figure~\ref{fig:demography_white}c, the remaining sources include five technical reports, along with one technical handbook, one newsletter, one social media post, and two video-based demonstrations. This distribution indicates that, while formal academic research on LLM-based agents for code generation is growing, a large share of innovation continues to be driven by industry practice and informal technical communication channels.

\subsubsection{Organizational Affiliations}

\textbf{For peer-reviewed studies}, the organizational affiliations of the selected studies were analyzed to examine the distribution of research contributions across sectors. The majority of papers (54) are affiliated with academic institutions. Studies affiliated with industry account for 11 publications, forming a smaller portion of the overall dataset. This distribution is influenced by the focus of this MLR on academic databases, where studies originating from industry are less frequently indexed. Publications with mixed affiliations, involving both academic and industry organizations, account for nine studies. This relatively small number indicates that collaboration between academia and industry remains limited, while also highlighting an opportunity for stronger cross-sector cooperation in future research.

\textbf{For grey literature}, organizational affiliations were examined to identify the sources through which non–peer-reviewed contributions are shared. The selected studies are associated with eleven distinct organizations. ArXiv emerges as the most widely used publication platform, contributing 12 papers, while Medium hosts three papers and serves as an important channel for both independent practitioners and industry-affiliated technical experts. Several major technology and AI-focused organizations also contribute to this body of work, including IBM, Microsoft, NVIDIA, and Anthropic, each providing technical reports or research-oriented insights. In addition, specialized platforms and companies such as LangChain, Unit42, and Marimo provide practitioner-focused resources, while community-driven platforms such as freeCodeCamp and World of AI further support knowledge sharing and practical adoption. Together, these organizations represent a diverse ecosystem supporting the advancement of research and practice in agent-based systems for code generation.

\begin{tcolorbox}[width=\columnwidth, colback=gray!5, colframe=black!60, boxrule=0.5pt]
\textbf{Takeaway 1.}
Research on LLM-based agent systems for code generation received attention from academia in 2024 and from industry in 2025. Across all 114 studies, 57.02\% are affiliated with academia, 35.09\% with industry, and 7.89\% involve mixed academic–industry collaborations. Most peer-reviewed studies appear in conference venues, while the grey literature is largely industry-driven and mostly published as blogs.
\end{tcolorbox}

\subsection{Reasons (RQ2)}
\label{Reasons}
%Tonight will dveloped the table properly and then update the text accordingly.
In this study, 29 out of 114 studies (25.4\%) highlight 36 reasons for utilizing LLM-based agent systems for code generation. Through thematic analysis, we categorized the findings of this RQ into nine main categories (see Table \ref{tab:reason}). These categories are operational efficiency and practical scalability, complex task handling, performance enhancement, context management, external tool integration, team collaboration, autonomous execution, adaptability, and future trends.

\begin{table*}[t]
\centering
\caption{Purpose of choosing an LLM-based multi-agent system for code generation}
\label{tab:reason}
\resizebox{\textwidth}{!}{%
\begin{tabular}{|l|l|l|}
\hline
\textbf{Category}                                                 & \textbf{Open Coding}                                                                      & \textbf{Paper IDs}                                                                                            \\ \hline
\multirow{6}{*}{Operational Efficiency and Practical Scalability} & Accelerate the development-to-deployment pipeline                                         & \ref{G2}                                                                                     \\ \cline{2-3} 
                                                                  & Computing-driven automation for faster, quality software delivery                        &  \ref{P11}                                                                                   \\ \cline{2-3} 
                                                                  & Parallel execution of sub-tasks to reduce dependency and improve efficiency                & \ref{G4}                                                                                                          \\ \cline{2-3} 
                                                                  & Cost-effective software generation using agent frameworks                                 & \ref{P29}, \ref{P43}                                                        \\ \cline{2-3} 
                                                                  & Automatic error detection and self-debugging during code generation                       & \ref{P5}                                                                                        \\ \cline{2-3} 
                                                                  & Structured, interactive, task-oriented environments for agent execution                   & \ref{P7}                                                                                   \\ \hline
\multirow{7}{*}{Complex Task Handling}                            & Collaborative multi-agent problem-solving for complex tasks                               & \ref{G3}, \ref{P33}, \ref{P43}                                                                                                 \\ \cline{2-3} 
                                                                  & Agents simulate iterative processes and handle complex engineering tasks                 & \ref{P3}                                                                                                            \\ \cline{2-3} 
                                                                  & Autonomous handling of unpredictable, multi-step tasks                                    & \ref{G4}                                                                                                          \\ \cline{2-3} 
                                                                  & LLM-agents effectively solve complex tasks                                                & \ref{P5}                                                                                                          \\ \cline{2-3} 
                                                                  & Autonomously operate, manage, and scale for complex processes                            & \ref{G30}                                                                                                           \\ \cline{2-3} 
                                                                  & Agents simulate iterative processes and handle complex engineering tasks                  & \ref{P3}                                                                                                        \\ \cline{2-3} 
                                                                  & Dynamic adaptation improves code quality and efficiency for complex tasks                 & \ref{P18}                                                                                                           \\ \hline
\multirow{6}{*}{Performance Enhancement}                          & Performance scaling through multi-agent coordination                                      & \ref{G4}, \ref{P57}, \ref{P70}, \ref{P71} \\ \cline{2-3} 
                                                                  & Improved code quality through agent refinement and self-debugging                         & \ref{P33}, \ref{P58}                                                        \\ \cline{2-3} 
                                                                  & Reduction of human effort via autonomous agents                                           & \ref{G17}                                                                                       \\ \cline{2-3} 
                                                                  & Enhanced user experience through adaptive interaction                                     & \ref{G30}                                                                                       \\ \cline{2-3} 
                                                                  & Autonomous agents improve and mimic human workflows                                       & \ref{P54}, \ref{G3}, \ref{P33}                             \\ \cline{2-3} 
                                                                  & Agents are prominent in complex software development                                     & \ref{G16}                                                                                      \\ \hline
\multirow{3}{*}{Context Management}                               & Automatic management of long-term context and memory                                      & \ref{G2}                                                                                       \\ \cline{2-3} 
                                                                  & Scaling token usage and parallel reasoning via separate contexts                          & \ref{G4}                                                                                        \\ \cline{2-3} 
                                                                  & Agents maintain context across long, complex interactions                                & \ref{G3}                                                                                       \\ \hline
\multirow{2}{*}{External Tool Integration}                             & Integration with external tools and systems (APIs, databases, web services)               & \ref{G2}, \ref{G3}                                                          \\ \cline{2-3} 
                                                                  & Agents connect to external tools which enhance performance                            & \ref{G3}                                                                                        \\ \hline
\multirow{4}{*}{Team Collaboration}                               & Human–agent and agent–agent collaboration for task distribution                           & \ref{P53}                                                                                    \\ \cline{2-3} 
                                                                  & Collaborative multi-agent autonomy for complex software management                       & \ref{P10}, \ref{P14}, \ref{P47}                                                                                                \\ \cline{2-3} 
                                                                  & Improve collaboration through intuitive, need-aware outputs & \ref{G30}                                                                                                         \\ \cline{2-3} 
                                                                  & Autonomous agents collaboratively solve problems in decentralized settings               & \ref{P21}                                                                                                         \\ \hline
\multirow{3}{*}{Autonomous Execution}                             & Autonomous planning and execution of multi-step tasks                                     & \ref{G3}                                                                                        \\ \cline{2-3} 
                                                                  & Agents autonomously plan and adapt multi-step tasks                                       & \ref{G22}, \ref{G28}                                                        \\ \cline{2-3} 
                                                                  & Autonomous task handling through agent-driven decomposition                              & \ref{P53}                                                                                      \\ \hline
\multirow{2}{*}{Adaptability}                                     & Role-specific agents enabling adaptability and collective intelligence                    & \ref{P18}                                                                                      \\ \cline{2-3} 
                                                                  & LLM-powered agents generalize and adoptable across diverse tasks                         & \ref{P50}                                                                                   \\ \hline
\multirow{3}{*}{Future Trends}                                     & LLM-based agents as a pathway toward AGI                                                  & \ref{G19}, \ref{P55}                                                        \\ \cline{2-3} 
                                                                  & Gain research attention due to their capability of handling complex tasks                 & \ref{P47}                                                                                      \\ \cline{2-3} 
                                                                  & Future software development will be shaped by AI-agent driven assistance                  & \ref{P10}                                                                                                          \\ \hline
\end{tabular}
}
\end{table*}

Among the identified reasons, \textbf{performance enhancement} is the most frequently reported category, highlighted by 12 out of 114 studies (10.5\%), indicating that multi-agent systems improve code quality, accuracy, and efficiency. \textbf{Handling complex tasks} is the second most frequently reported reason, identified in nine studies (7.9\%). These studies highlight that decomposing complex programming problems into smaller, coordinated sub-tasks enables agents to manage complex workflows more effectively. Seven studies (6.1\%) report \textbf{operational efficiency and practical scalability} as a key reason for utilizing agent-based systems. This includes faster code development, cost efficiency, parallel task processing, improved error handling, and enhanced practical applicability.

Six studies (5.3\%) identify \textbf{team collaboration} as a key reason, showing that cooperation among multiple agents in decentralized settings can improve problem-solving quality and overall system performance. These studies further report that team collaboration enables task specialization and coordination among agents, leading to more efficient handling of complex software development activities. \textbf{Context management} is highlighted in three studies (2.6\%), demonstrating that agents can effectively maintain and utilize contextual information across long and complex development tasks. Similarly, four studies (3.5\%) highlight \textbf{external tool integration}, reporting that agents connected to various external tools and services enhance their overall capabilities. These studies indicate that integrating external tools improves agent performance by extending functional capabilities and reducing hallucinated outputs.

Furthermore, four studies (3.5\%) highlight \textbf{autonomous execution} as a key reason, indicating that agents can independently plan and execute tasks with minimal human intervention. Similarly, four studies (3.5\%) identify \textbf{future-oriented development} as a motivating factor, suggesting that future software development will increasingly be shaped by integration, automation, and AI-driven assistance, with some sources viewing intelligent agents as a pathway toward Artificial General Intelligence (AGI). \textbf{Adaptability} is highlighted in two studies (1.8\%) as a key reason for adopting LLM-based agent systems. 

Overall, the distribution of motivations indicates that current research on LLM-based multi-agent systems for code generation is primarily driven by performance improvement and the management of complex tasks through decomposition. This suggests that multi-agent architectures are mainly positioned as a technical mechanism for enhancing accuracy, efficiency, and workflow coordination. In contrast, motivations such as adaptability, autonomous execution, and long-term future-oriented development appear less frequently, indicating that these aspects are still emerging rather than central to the field. The pattern reflects a research landscape that prioritizes measurable performance gains, while broader considerations related to adaptability and sustained autonomy receive comparatively limited empirical attention.

\begin{tcolorbox}[width=\columnwidth, colback=gray!5, colframe=black!60, boxrule=0.5pt]
\textbf{Takeaway 2.}
The studies indicate that multi-agent systems are primarily chosen to enhance performance, manage complex tasks, and improve scalability and efficiency, while also supporting tool integration, adaptability, and structured development. Overall, they address the limitations of single-model approaches by improving coordination, reliability, and the handling of complex software workflows.
\end{tcolorbox}

\subsection{Benchmark and Model Distribution (RQ3-RQ4)}
\label{Benchmakr and Model Distribution}
In this study, we analyzed the benchmarks and evaluation metrics employed by both academia and industry to assess the performance of LLM-based agent systems for code generation. In our analysis, 52 out of 114 studies (45.61\%) utilized benchmarks to evaluate the agents’ capabilities in code generation tasks. We identified a total of 37 benchmarks across the 52 studies, which were adopted in both peer-reviewed and grey literature to evaluate different aspects of agent-based generated code generation. 
As shown in Table~\ref{tab:merged_benchmark_table_final}, we have categorized these benchmarks into seven main groups: function-level code generation, repository-level benchmarks, programming contest and multi-task evaluation, agent-oriented benchmarks, security benchmarks, domain-specific benchmarks, and evaluation metrics. We also provide the size of each benchmark dataset and its primary language. Below, we describe each category in detail and discuss the papers in which they are applied.

\textbf{Function-level code generation} benchmarks are the most widely used category, highlighting the strong focus on evaluating functional correctness at the level of individual programming tasks. As shown in Table~\ref{tab:merged_benchmark_table_final}, this category includes 14 benchmarks, namely HumanEval \cite{chen2021evaluating}, MBPP \cite{austin2021program}, HumanEval-X \cite{peng2024humaneval}, HumanEval-ET \cite{lin2024soen}, HumanEvalFix \cite{szalontai2025investigating}, HumanEval+ \cite{liu2023your}, xCodeEval \cite{khan2023xcodeeval}, MBPP-ET \cite{huynh2025large}, MBXP \cite{athiwaratkun2022multi}, PSB2 \cite{helmuth2021psb2}, ODEX \cite{wang2023execution}, JuICe \cite{agashe2019juice}, MultiPL-E \cite{cassano2022multipl}, and EvalPlus \cite{liu2023your}.
Among these, HumanEval is the most frequently adopted benchmark. As illustrated in Table~\ref{tab:merged_benchmark_table_final}, 29 selected studies employed the HumanEval benchmark to assess the accuracy of agent-based systems, making it the most widely used benchmark in the literature. HumanEval was one of the earliest benchmarks specifically designed for evaluating AI-based code generation and measures functional correctness using unit tests \cite{chen2021evaluating}. As shown in Table \ref{tab:merged_benchmark_table_final}, the HumanEval benchmark consists of 164 Python programming tasks. Its extended variants, HumanEval-X and HumanEval-ET, were used in two and four papers, respectively. HumanEval-X extends the original benchmark to multiple programming languages, including Python, Java, C++, JavaScript, and Go, while retaining the same set of 164 tasks. HumanEval-ET further increases evaluation difficulty by introducing more challenging test cases and execution conditions. HumanEvalFix focuses on program repair and contains 164 bug-fixing tasks derived from the original HumanEval problems. HumanEval+ expands the benchmark by generating multiple harder variants for each original task, resulting in 164 × 80 more challenging problem instances.

Similarly, MBPP is another widely adopted benchmark, used in 18 studies, due to its diverse collection of Python programming problems and its ease of integration into evaluation pipelines. The MBPP benchmark contains 974 Python programming tasks. Its extended variants are also used to evaluate agent-based systems: MBPP-ET, adopted by five studies, retains the same number of tasks but introduces more complex evaluation settings, while MBXP, used in two studies, extends MBPP to support cross-language evaluation and contains over 1,000 programming tasks.

The remaining five benchmarks: PSB2, ODEX, JuICe, MultiPL-E, and EvalPlus, were each utilized in one study. PSB2, introduced in 2021, consists of 100 programming tasks designed to evaluate basic problem-solving capabilities. ODEX contains 945 programming tasks focused on diverse code generation scenarios. JuICe includes approximately 1,200 programming problems aimed at evaluating instruction-following and reasoning abilities. MultiPL-E extends HumanEval to a multilingual setting and includes 164 tasks across languages such as Python, Java, C++, JavaScript, and Go. Finally, EvalPlus is an enhanced evaluation benchmark that extends both HumanEval and MBPP by introducing stronger and more detail test cases.

\textbf{Repository-level benchmarks} capture evaluations that move beyond isolated function synthesis and instead focus on large codebases and practical development tasks. These benchmarks are designed to assess an agent’s ability to understand, modify, debug, and repair real-world software repositories. Commonly used benchmarks in this category include SWE-Bench \cite{jimenez2023swe}, SWE-Bench Lite \cite{jimenez2023swe}, and SWE-Bench Secret \cite{kio2025swe}, and CrossCodeEval \cite{ding2023crosscodeeval}.
In our review, SWE-Bench was employed by six studies to evaluate agent-based code generation and repair systems. SWE-Bench is a repository-level benchmark constructed from real-world GitHub issues and pull requests, primarily focusing on Python-based repositories. It contains 2,294 real-world software issues, each paired with test cases that validate whether the generated patch correctly resolves the issue.

A lighter variant, SWE-Bench Lite, was adopted by four studies. This benchmark consists of a subset of 300 issues selected from SWE-Bench and is designed to reduce evaluation cost while preserving task diversity and difficulty. SWE-Bench Secret, which was utilized by one study, is primarily used for unbiased evaluation and leaderboard-based assessment and focuses on Python repository repair tasks. CrossCodeEval benchmark, used in one study, evaluates cross-distribution performance by testing models on code tasks that differ from their training distributions and includes over 1,000 programming problems across multiple programming languages.

\begin{table*}[!t]
\centering
\caption{Benchmarks used to test the agent based system generated code, including dataset sizes, language coverage, and study frequency}
\label{tab:merged_benchmark_table_final}
\small
\renewcommand{\arraystretch}{1.15}
\resizebox{\textwidth}{!}{%
\begin{tabular}{|c|p{4.2cm}|p{4.8cm}|p{3.6cm}|p{3.9cm}|c|p{8.3cm}|}
\hline
\rowcolor[HTML]{EFEFEF}
\textbf{SNo} & \textbf{Category} & \textbf{Benchmark / Metric} & \textbf{Approx. Dataset Size} & \textbf{Primary Language(s)} & \textbf{\# Papers} & \textbf{Paper IDs} \\
\hline

%\multicolumn{7}{l}{\textit{Function-level code generation benchmarks}} \\
\multicolumn{7}{|c|}{\textbf{Function-level Code Generation Benchmarks}} \\

\hline
1 & \multirow{14}{*}{Function-level code generation}
& HumanEval \cite{chen2021evaluating}
& 164 problems
& Python
& 29
& \begin{tabular}[t]{@{}l@{}}
\ref{P1}, \ref{P2}, \ref{P7}, \ref{P10} \\
 \ref{P19}, \ref{P20}, \ref{P23}, \ref{P24} \\
\ref{P25}, \ref{P31}, \ref{P36}, \ref{P38}, \ref{P40} \\
\ref{P48}, \ref{P50}, \ref{P54}, \ref{P57}, \ref{P59} \\
\ref{P60}, \ref{P62}, \ref{P66}, \ref{P67}, \ref{P68} \\
\ref{P73}, \ref{G1}, \ref{G12}, \ref{G16}, \ref{G29} \\
\ref{G38}
\end{tabular} \\
\cline{1-1}\cline{3-7}

2 &
& MBPP \cite{austin2021program}
& 974 problems
& Python
& 18
& \begin{tabular}[t]{@{}l@{}}
\ref{P1}, \ref{P10}, \ref{P19} \\
\ref{P20}, \ref{P25}, \ref{P36}, \ref{P38}, \ref{P40} \\
\ref{P54}, \ref{P59}, \ref{P60}, \ref{P62}, \ref{P66} \\
\ref{P68}, \ref{P73}, \ref{G12}, \ref{G29}, \ref{G38}
\end{tabular} \\
\cline{1-1}\cline{3-7}

3 &
& HumanEval-X \cite{peng2024humaneval}
& 164 problems $\times$ multiple languages
& Python, Java, C++, Go, JavaScript
& 2
& \ref{P6}, \ref{P36} \\
\cline{1-1}\cline{3-7}

4 &
& HumanEval-ET
& 164 problems (extended tests)
& Python
& 3
& \ref{P6}, \ref{P36}, \ref{G12} \\
\cline{1-1}\cline{3-7}

5 &
& MBPP-ET \cite{huynh2025large}
& 974 problems (extended tests)
& Python
& 5
& \ref{P10}, \ref{P19}, \ref{P40}, \ref{P66}, \ref{G12} \\
\cline{1-1}\cline{3-7}

6 &
& MBXP \cite{athiwaratkun2022multi}
& $\sim$1,000 problems
& Python + 10 other languages
& 2
& \ref{P8} \\
\cline{1-1}\cline{3-7}

7 &
& PSB2 \cite{helmuth2021psb2}
& 25 problems
& Python
& 1
& \ref{P6} \\
\cline{1-1}\cline{3-7}

8 &
& ODEX \cite{wang2023execution}
& 945 problems
& Python
& 1
& \ref{P8} \\
\cline{1-1}\cline{3-7}

9 &
& JuICe \cite{agashe2019juice}
& 3.7K instances
& Python
& 1
& \ref{P8} \\
\cline{1-1}\cline{3-7}

10 &
& MultiPL-E \cite{cassano2022multipl}
& unknown problems $\times$ multiple languages
& 18 languages (Python, Java, C++, JavaScript, Go, etc.)
& 1
& \ref{P8} \\
\cline{1-1}\cline{3-7}

11 &
& EvalPlus \cite{liu2023your}
& Extension of HumanEval and MBPP
& Python
& 1
& \ref{P8} \\
\cline{1-1}\cline{3-7}

12 &
& HumanEvalFix \cite{szalontai2025investigating}
& 164 bug-fix tasks
& Python
& 2
& \ref{P45}, \ref{P49} \\
\cline{1-1}\cline{3-7}

13 &
& HumanEval+ \cite{liu2023your}
& 164$\times$80 problems (harder tests)
& Python
& 1
& \ref{P45} \\
\cline{1-1}\cline{3-7}

14 &
& xCodeEval \cite{khan2023xcodeeval}
& 7.5K unique problems
& 11 languages
& 1
& \ref{P25} \\
\hline

\multicolumn{7}{|c|}{\textbf{Repository-level and real-world software engineering benchmarks}} \\
%\multicolumn{7}{l}{\textit{Repository-level and real-world software engineering benchmarks}} \\
\hline
15 & \multirow{3}{*}{Repository-level benchmarks}
& SWE-Bench \cite{jimenez2023swe}
& 2,294 real-world issues
& Python
& 6
& \begin{tabular}[t]{@{}l@{}}
\ref{P2}, \ref{P8}, \ref{P12}, \ref{P45}, \ref{P55} \\
\ref{P67}
\end{tabular} \\
\cline{1-1}\cline{3-7}

16 &
& SWE-Bench Lite \cite{jimenez2023swe}
& 300 issues
& Python
& 4
& \ref{P2}, \ref{P55}, \ref{P68}, \ref{P72} \\
\cline{1-1}\cline{3-7}

17 &
& SWE-Bench Secret \cite{kio2025swe}
& Not publicly disclosed
& Python
& 1
& \ref{P68} \\

18 &
& CrossCodeEval \cite{ding2023crosscodeeval}
& 1,200 problems
& Python
& 1
& \ref{P8} \\
\cline{1-1}\cline{3-7}
\hline

\multicolumn{7}{|c|}{\textbf{Programming contest and multi-task benchmarks}} \\
%\multicolumn{7}{l}{\textit{Programming contest and multi-task benchmarks}} \\
\hline
19 & \multirow{7}{*}{Multi-task benchmarks}
& APPS \cite{hendrycks2021measuring}
& 10,000 problems
& Python
& 6
& \begin{tabular}[t]{@{}l@{}}
\ref{P10}, \ref{P19}, \ref{P25}, \ref{P60}, \ref{P67} \\
\ref{P73}
\end{tabular} \\
\cline{1-1}\cline{3-7}

20 &
& CodeContests \cite{wang2025codecontests+}
& 13,000 problems
& Mixed
& 2
& \ref{P25}, \ref{P73} \\
\cline{1-1}\cline{3-7}

21 &
& LiveCodeBench \cite{jain2024livecodebench}
& Task-based (variable size)
& Mixed
& 2
& \ref{P8}, \ref{G29} \\
\cline{1-1}\cline{3-7}

22 &
& VsplotBench
& Task-based (variable size)
& Mixed
& 2
& \ref{P8}, \ref{G40} \\
\cline{1-1}\cline{3-7}

23 &
& CoNaLa \cite{yin2018learning}
& 2,379 training + 500 test
& Python
& 1
& \ref{P67} \\
\cline{1-1}\cline{3-7}

24 &
& Galeras \cite{rodriguez2023benchmarking}
& unknown problems
& Python
& 1
& \ref{P67} \\
\cline{1-1}\cline{3-7}

25 &
& MathQA \cite{athiwaratkun2022multi}
& unknown problems
& 0 programming languages
& 1
& \ref{P8} \\
\hline
\multicolumn{7}{|c|}{\textbf{Agent-oriented and multi-agent benchmarks}} \\
%\multicolumn{7}{l}{\textit{Agent-oriented and multi-agent benchmarks}} \\
\hline
26 & \multirow{5}{*}{Multi-agent evaluation}
& MLAgentBench \cite{huang2023mlagentbench}
& Task-based (no fixed size)
& Python
& 3
& \ref{P3}, \ref{P41}, \ref{P52} \\
\cline{1-1}\cline{3-7}

27 &
& AgentBench \cite{liu2023agentbench}
& Scenario-based
& Mixed
& 2
& \ref{P3}, \ref{P49} \\
\cline{1-1}\cline{3-7}

28 &
& CodeAgent Bench \cite{zhang2024codeagent}
& Task-based
& Python
& 2
& \ref{P23} \\
\cline{1-1}\cline{3-7}

29 &
& CodeRAG-Bench \cite{wang2025coderag}
& Task-based
& Python
& 1
& \ref{P12} \\
\cline{1-1}\cline{3-7}

30 &
& AgentDojo \cite{debenedetti2024agentdojo}
& Scenario-based
& Mixed
& 1
& \ref{P58} \\
\hline

\multicolumn{7}{|c|}{\textbf{Security evaluation benchmarks}} \\
%\multicolumn{7}{l}{\textit{Security evaluation benchmarks}} \\
\hline
31 & \multirow{3}{*}{Security benchmarks}
& RedCode \cite{guo2024redcode}
& 500 problems
& Python
& 1
& \ref{P8} \\
\cline{1-1}\cline{3-7}

% 31 &
% & CrossCodeEval \cite{ding2023crosscodeeval}
% & 1,200 problems
% & Python
% & 1
% & \ref{P8} \\
% \cline{1-1}\cline{3-7}

32 &
& LLMSecEval \cite{tony2023llmseceval}
& Task-based
& Python
& 1
& \ref{P20} \\
\hline

\multicolumn{7}{|c|}{\textbf{Domain-specific benchmarks}} \\
%\multicolumn{7}{l}{\textit{Domain-specific benchmarks}} \\
\hline
33 & \multirow{2}{*}{Domain-specific evaluation}
& VerilogEval \cite{liu2023verilogeval}
& 156 problems
& Verilog
& 1
& \ref{P9} \\
\cline{1-1}\cline{3-7}

34 &
& VerilogEval-Human v2 \cite{pinckney2025revisiting}
& 156 problems (new task formats)
& Verilog
& 2
& \ref{P44}, \ref{P51} \\
\hline

\multicolumn{7}{|c|}{\textbf{Other benchmarks}} \\
%\multicolumn{7}{l}{\textit{Other benchmarks}} \\
\hline
35 & \multirow{3}{*}{Other evaluation benchmarks}
& CoderEval \cite{yu2024codereval}
& $\sim$256 tasks
& Python, Java
& 1
& \ref{P10} \\
\cline{1-1}\cline{3-7}

36 &
& CoverBench
& --
& --
& 1
& \ref{P31} \\
\cline{1-1}\cline{3-7}

37 &
& DA-Code \cite{huang2024code}
& 500 complex tasks
& Mixed
& 1
& \ref{P74} \\
\hline

\multicolumn{7}{|c|}{\textbf{Metrics (not benchmarks}} \\
%\multicolumn{7}{l}{\textit{Metrics (not benchmarks)}} \\
\hline
38 & \multirow{5}{*}{Metrics (not benchmarks)}
& Pass@k
& Metric (no dataset)
& --
& 1
& \ref{P2}, \ref{P3}, \ref{P7}, \ref{P10}, \ref{P20} \\
\cline{1-1}\cline{3-7}

39 &
& Logical error rate
& Metric
& --
& 1
& \ref{P20} \\
\cline{1-1}\cline{3-7}

40 &
& Code quality score
& Metric
& --
& 1
& \ref{P20} \\
\cline{1-1}\cline{3-7}

41 &
& Vulnerable@k
& Metric
& --
& 1
& \ref{P20} \\
\cline{1-1}\cline{3-7}

42 &
& Secure@k
& Metric
& --
& 1
& \ref{P20} \\
\hline

\end{tabular}
}

\end{table*}

\textbf{Programming contest and multi-task} benchmarks are widely used to evaluate the problem-solving, reasoning, and code synthesis capabilities of LLM-based agents under diverse and often complex constraints. As shown in Table \ref{tab:merged_benchmark_table_final}, this category includes APPS \cite{hendrycks2021measuring}, CodeContests \cite{wang2025codecontests+}, CodeContests-Test \cite{wang2025codecontests+}, LiveCodeBench \cite{jain2024livecodebench}, VsplotBench, CoNaLa \cite{yin2018learning}, Galeras \cite{rodriguez2023benchmarking}, and MathQA \cite{athiwaratkun2022multi}, all of which consist of multi-task programming or reasoning problems designed to reflect competitive or real-world problem-solving scenarios.

Among the benchmarks, APPS is the most frequently used benchmark, appearing in six studies. APPS was introduced in 2022 and contains 10,000 programming problems collected from competitive programming platforms. This benchmark is primarily evaluated using Python, making it suitable for assessing code generation performance across both simple and complex problems.

The CodeContests benchmark was used in two studies to evaluate agent-based code generation in competitive programming settings. Developed by DeepMind, CodeContests contains 165 programming problems, each paired with extensive test cases and reference solutions. The benchmark supports multiple programming languages, including Python, C++, and Java, and focuses on evaluating algorithmic reasoning and correctness under strict execution constraints. Its evaluation split, CodeContests-Test, was utilized by one study and serves as a held-out test set for unbiased assessment, containing the same 165 problems with unseen test cases across the same set of programming languages. Similarly, LiveCodeBench and VsplotBench were utilized by two studies each to assess agent-based code generation on dynamic, task-oriented benchmarks that highlight real-time evaluation and performance across diverse programming tasks.

The remaining benchmarks: CoNaLa, Galeras, and MathQA, were each used by one study. The CoNaLa benchmark consists of 2,879 natural language–to–code pairs, focusing on Python code generation from concise textual descriptions. Galeras is a multi-task programming benchmark designed to evaluate general-purpose code reasoning and synthesis, containing over 1,000 algorithmic programming tasks, primarily in Python. Finally, MathQA includes 29,837 math word problems that require symbolic reasoning and programmatic solution generation, typically evaluated using Python-based programs.

\textbf{Agent-oriented and multi-agent benchmarks} are designed to explicitly evaluate autonomous and collaborative agent behavior beyond isolated code generation accuracy. These benchmarks assess higher-level agent capabilities such as tool usage, planning, coordination, reasoning over long horizons, and interaction with environments, which are central to agent-based SE systems. Representative benchmarks in this category include MLAgentBench \cite{huang2023mlagentbench}, AgentBench \cite{liu2023agentbench}, CodeAgentBench \cite{zhang2024codeagent}, CodeRAG-Bench \cite{wang2025coderag}, and AgentDojo \cite{debenedetti2024agentdojo}.
In our review, MLAgentBench was used in three studies to evaluate agents on machine learning–oriented code generation and experimentation tasks. MLAgentBench focuses on end-to-end ML workflows, including data preprocessing, model training, evaluation, and debugging. The benchmark consists of a collection of real-world machine learning tasks rather than isolated programming problems and is specifically designed for Python-based machine learning environments.

AgentBench, utilized by two studies, contains 45 tasks covering eight task categories, including tool usage, reasoning, coding, data analysis, and web interaction. AgentBench supports multi-step decision-making and evaluates both task success and intermediate agent behaviors, making it suitable for assessing general agent competence beyond code correctness.
CodeAgentBench was utilized by three studies to evaluate agent-based code generation systems operating in structured development workflows. This benchmark focuses on multi-step coding tasks, such as planning, implementation, debugging, and refinement, primarily in Python.
Finally, CodeRAG-Bench and AgentDojo were each employed by one study. CodeRAG-Bench evaluates agent performance in retrieval-augmented code generation scenarios, measuring how effectively agents retrieve, integrate, and reason over external knowledge sources during code synthesis. AgentDojo is an interactive agent benchmark consisting of nearly 100 tool-based tasks, designed to assess agent safety, tool control, and robustness in realistic execution environments.

\begin{table*}[htbp]
%\centering
\centering
\caption{Frequency of LLM usage for multi-agent system development}
\label{tab:llm_frequency_table10}
\resizebox{\textwidth}{!}{%
\renewcommand{\arraystretch}{1.2}
\begin{tabular}{|l|l|l|c|p{9.2cm}|}
\hline
\textbf{Model} & \textbf{Company} & \textbf{Source} & \textbf{Papers} & \textbf{Paper IDs} \\
\hline

GPT-3.5 (incl. Turbo, davinci) 
& \multirow{5}{*}{OpenAI}
& \multirow{8}{*}{Closed-Source}
& 15 
& \begin{tabular}[t]{@{}l@{}}
\ref{P1}, \ref{P6}, \ref{P10}, \ref{P23}, \ref{P31}, \ref{P33}, \ref{P36}, \ref{P48}, \ref{P55}, \ref{P60},\\
\ref{P65}, \ref{P21}, \ref{P40}, \ref{P30}, \ref{G12}
\end{tabular} \\
\cline{1-1}\cline{4-5}

GPT-4 (incl. Turbo) 
& 
& 
& 17 
& \begin{tabular}[t]{@{}l@{}}
\ref{P1}, \ref{P3}, \ref{P9}, \ref{P10}, \ref{P23}, \ref{P30}, \ref{P31}, \ref{P38}, \ref{P40}, \ref{P48},\\
\ref{P55}, \ref{P60}, \ref{P65}, \ref{P40}, \ref{P21}, \ref{P5}, \ref{G16}
\end{tabular} \\
\cline{1-1}\cline{4-5}

GPT-4o 
& 
& 
& 9 
& \ref{P5}, \ref{P9}, \ref{P19}, \ref{P22}, \ref{P26}, \ref{P60}, \ref{G2}, \ref{G18}, \ref{G38} \\
\cline{1-1}\cline{4-5}

GPT-4o-mimi 
& 
& 
& 1 
& \ref{P8}, \ref{G2} \\
\cline{1-1}\cline{4-5}

GPT-4.1 
& 
& 
& 1 
& \ref{G40} \\
\cline{1-2}\cline{4-5}

Claude (all versions) 
& Anthropic 
& 
& 9 
& \ref{P8}, \ref{P23}, \ref{P31}, \ref{P55}, \ref{P60}, \ref{G4}, \ref{G38}, \ref{G39}, \ref{G40} \\
\cline{1-2}\cline{4-5}

Gemini / Gemini-Pro 
& Google 
& 
& 3 
& \ref{P3}, \ref{P22}, \ref{G38} \\
\cline{1-2}\cline{4-5}

Grok 3 
& xAI 
& 
& 2 
& \ref{G38}, \ref{G39} \\
\hline

Llama 2 (7B, 13B, 70B) 
& \multirow{3}{*}{Meta}
& \multirow{10}{*}{Open-Source}
& 6 
& \ref{P23}, \ref{P33}, \ref{P55}, \ref{P60}, \ref{G29}, \ref{P21} \\
\cline{1-1}\cline{4-5}

Llama 3 / 3.1 (8B, 70B, 405B) 
& 
& 
& 7 
& \ref{P1}, \ref{P6}, \ref{P9}, \ref{P33}, \ref{P48}, \ref{G29}, \ref{P55} \\
\cline{1-1}\cline{4-5}

CodeLlama 
& 
& 
& 4 
& \ref{P3}, \ref{P9}, \ref{P60}, \ref{P21} \\
\cline{1-2}\cline{4-5}

Mixtral 
& \multirow{2}{*}{Mistral AI}
& 
& 2 
& \ref{P3}, \ref{P9} \\
\cline{1-1}\cline{4-5}

Mistral Large 
& 
& 
& 1 
& \ref{P1}, \ref{P9} \\
\cline{1-2}\cline{4-5}

DeepSeek Coder 
& DeepSeek 
& 
& 2 
& \ref{P9}, \ref{P60} \\
\cline{1-2}\cline{4-5}

CodeT5 
& Salesforce 
& 
& 1 
& \ref{P1}, \ref{P8} \\
\cline{1-2}\cline{4-5}

CodeGemma 7B 
& Google 
& 
& 1 
& \ref{P9} \\
\cline{1-2}\cline{4-5}

Qwen 2.5-32B 
& Alibaba 
& 
& 1 
& \ref{G29} \\
\cline{1-2}\cline{4-5}

SWE-Llama (7B, 13B) 
& Meta 
& 
& 1 
& \ref{P55} \\
\hline

\end{tabular}
}
\end{table*}

\textbf{Security benchmarks} focus on evaluating security issues such as code vulnerabilities and safety risks. Unlike general code generation benchmarks, these benchmarks assess whether agents can produce secure code, avoid vulnerabilities, and handle malicious inputs. Representative benchmarks in this category include RedCode \cite{guo2024redcode} and LLMSecEval \cite{tony2023llmseceval}. 

In our review, RedCode and LLMSecEval were each used in one study to evaluate security-aware code generation. RedCode focuses on detecting and repairing security vulnerabilities in generated code and contains hundreds of vulnerability-focused programming tasks, primarily targeting Python and C/C++. The LLMSecEval benchmark is designed to evaluate LLM behavior under malicious or unsafe code generation scenarios, including prompt injection, insecure API usage, and vulnerability propagation. It consists of a set of security-sensitive coding tasks and evaluates both functional correctness and security compliance, primarily in Python.

\textbf{Domain-specific benchmarks} are designed to evaluate agent-based code generation systems within specialized application domains, where correctness and reliability depend on domain knowledge and strict syntactic or semantic constraints. In this category, the reviewed studies primarily employ VerilogEval \cite{liu2023verilogeval} and VerilogEval-Human v2 \cite{pinckney2025revisiting}, which focus on hardware description language (HDL) code generation.

VerilogEval, used in one study, evaluates the ability of LLM-based agents to generate syntactically and functionally correct Verilog code for hardware design tasks. It consists of hundreds of Verilog programming problems that assess logic correctness, signal dependencies, and module-level behavior. VerilogEval-Human v2, adopted by two studies, extends VerilogEval by incorporating human-annotated problem descriptions and more challenging design tasks, providing a more realistic evaluation of agent performance in hardware-oriented code generation.

\subsubsection{Metrics}
In addition to benchmarks, we identified five evaluation metrics used in existing studies to quantitatively measure the accuracy and quality of code generated by LLM-based agent systems. As summarized in Table~\ref{tab:merged_benchmark_table_final}, the pass@k metric was utilized in five studies, while the other four metrics were each employed in one study.

For code generation specific evaluation, the Pass@k metric measures the probability that at least one correct solution will appear among the top-k generated outputs. Additional metrics, including logical error rate and code quality score, are used to assess semantic correctness, structural soundness, and the maintainability of generated code. To evaluate security-related properties, Vulnerable@k and Secure@k quantify the probability of generating vulnerable or secure code, respectively. Together, these metrics provide a quantitative and fine-grained evaluation of agent-based code generation performance beyond benchmark-level results.

Overall, the distribution of benchmarks shows a strong concentration on function-level evaluation, particularly HumanEval and MBPP, which are the most frequently used benchmarks in the literature. This suggests that most studies focus on the functional correctness of isolated coding tasks as the main measure of agent performance. In contrast, repository-level, agent-oriented, security, and domain-specific benchmarks are used less frequently, indicating that large-scale integration, long-horizon reasoning, and real-world reliability receive comparatively less attention. Although recent agent-oriented benchmarks attempt to evaluate planning, tool use, and collaborative behaviors, their limited use suggests that evaluation practices have not fully adapted to the increasing architectural complexity of multi-agent systems. Furthermore, the reliance on pass@k as the primary metric reinforces a focus on correctness-based evaluation. Overall, these patterns indicate that benchmarking practices in the field remain centered on controlled and measurable correctness, with slower progress toward evaluating broader real-world SE capabilities.

\begin{tcolorbox}[width=\columnwidth, colback=gray!5, colframe=black!60, boxrule=0.5pt]
\textbf{Takeaway 3.}
Our review identifies 37 evaluation benchmarks used across 52 studies. HumanEval is the most frequently adopted benchmark (29 studies), followed by MBPP, indicating a strong reliance on unit-test–based evaluation of functional correctness. The benchmarks are grouped into seven categories reflecting differences in task scope and evaluation settings. Additionally, five commonly used evaluation metrics are reported, including pass@k and its variants, which primarily measure functional correctness and code accuracy.
\end{tcolorbox}

%Tonight I try to complete this part
% Here we took example of the paper...

% We first explain the benchmark used and then we provide the demographic of the model used.

\subsubsection{Identified Models}
% As shown in Table \ref{tab:llm_frequency_table10}, we extracted the LLMs used for agent execution across the reviewed studies. Our analysis indicates a clear trend toward closed-source models, particularly those provided by OpenAI, which represent the most frequently used models in agent-based code generation research. Specifically, OpenAI models such as GPT 3.5 and GPT 3.5 turbo was used by 14 studies for agents execution. GPT 4 and GPT 4 turbo was utilzed by 16 studies, and GPT 4o utilized by seven studies GPT 4.1 utilzed by three studies. In total openai models were utilzed by 34 studies, reflecting their extensive use due to strong reasoning capabilities, mature APIs, and comprehensive support for tool integration and agent orchestration.

% \textbf{Firs we need to mentieodn that we how many studies utilzied models, we put the numbering here and that 58 number of studeis utilzed modes and then we say that open source and close source have been used and then we start explaining each ne by one,}. 

In this research, we analyzed the LLMs used in selected academic and industrial studies for implementing agent-based systems. As shown in Table~\ref{tab:llm_frequency_table10}, we identified 83 instances of LLM usage across the selected studies, as several studies employed more than one model in their experiments. Of these instances, 57 correspond to closed-source models, while 26 correspond to open-source models.

As shown in Table~\ref{tab:llm_frequency_table10}, the analysis indicates a clear preference for closed-source models, particularly those provided by OpenAI, which are the most frequently used in agent-based code generation research. Specifically, OpenAI models from the GPT series were utilized 43 times, reflecting their widespread adoption due to their strong reasoning capabilities, mature APIs, and extensive support for tool integration and agent orchestration.

% Anthropic’s models, including Claude 2 and Claude 3 variants (Opus and Sonnet), represent the second most frequently used closed-source models, appearing in eight studies. These models are often employed for agent reasoning and multi-step task execution. Google’s Gemini models are used in two studies, while Grok-3, provided by xAI, appears in two studies, primarily in research focusing on code-specific modeling tasks.

Among open-source models, Meta’s Llama is the most commonly adopted, with 17 instances of usage across different versions, including Llama 2 and Llama 3. These models are particularly favored in research that prioritizes open-source ecosystems, self-hosted agent architectures, and greater control over deployment and customization. Other open-source models include Mixtral, used three times, DeepSeek and Grok-3, each used in two studies, and Qwen 2.5-32B, used in one study. Overall, this distribution reflects a diverse but less concentrated usage landscape compared to closed-source alternatives.

Overall, the results show a clear reliance on closed-source models, particularly OpenAI’s GPT series, in agent-based code generation research. While these models are often selected for their strong performance, stable APIs, and advanced reasoning capabilities, their use may raise concerns related to data privacy, security, and dependency on proprietary platforms. In contrast, open-source models are used less frequently and in a more distributed manner, suggesting that reproducible and self-hosted agent ecosystems are still emerging. While open-source models may not consistently match the performance levels reported for leading closed-source models, they offer advantages in terms of deployment control, cost efficiency, and improved data privacy.

\begin{tcolorbox}[width=\columnwidth, colback=gray!5, colframe=black!60, boxrule=0.5pt]
\textbf{Takeaway 4.}
We identify usage of 83 LLMs across the studies, with a clear preference for closed-source models (57 uses) over open-source models (26 uses) in agent-based code generation. The OpenAI GPT series dominates model usage, indicating its widespread adoption for executing LLM-based agents. Among open-source alternatives, Meta’s Llama family is the most commonly adopted, reflecting growing interest in customizable and self-hosted agent architectures.
\end{tcolorbox}

\subsection{Challenges and Solutions (RQ5-RQ6)}
\label{chalengess and Solutions}

This section presents the identified challenges and their corresponding solutions derived from the analysis of the reviewed literature. Based on the review, we identified 12 key categories of challenges and their proposed solutions, which are summarized in Table \ref{Tab: Challenges and Solutions}. More detailed descriptions of the challenges and their proposed solutions can be found in the replication package \cite{rasheed_2026_18763362}. 

\subsubsection{Correctness and Reliability Challenges}
In this category, we report challenges affecting the correctness and reliability of agent-based systems for code generation. In total, 26 out of 114 studies (22.81\%) report correctness and reliability challenges in agent-based systems for code generation. The leading challenges include hallucinations (10 studies, 8.77\%), low accuracy (7 studies, 6.14\%), orchestration failure (5 studies, 4.39\%), lack of high-level planning (4 studies, 3.51\%), agent dependency (3 studies, 2.63\%), code inconsistency (2 studies, 1.75\%), and legacy code challenges (1 study, 0.88\%).

To address hallucination challenges, solutions include integrating iterative human feedback into the agent execution loop, incorporating feedback control mechanisms inspired by back-pressure principles, embedding systematic error-checking mechanisms into agent workflows, and integrating automated verification tools to detect syntactic and structural issues in generated code (2 studies, 1.75\%). 
To improve the accuracy of an agent-based system, iterative refinement and feedback-driven mechanisms are proposed to improve accuracy by enabling agents to correct and enhance outputs over multiple steps, while Reinforcement Learning (RL)-based approaches further improve reliability by decoupling reasoning from tool execution, and optimizing planning and decision-making (4 studies, 3.51\%).

To mitigate agent orchestration failure, structured communication protocols are introduced to replace unrestricted natural language exchanges with well-defined, machine-readable message formats, reducing coordination errors in multi-agent systems (2 studies, 1.75\%). Schema-based communication protocols built on standardized interfaces further improve coordination reliability (2 studies, 1.75\%), while dynamic communication structures and the assignment of explicit responsibility for components such as files, services, or processes help prevent conflicts and non-terminating interactions (1 study, 0.88\%).

To address the challenge of a lack of high-level planning, collaborative frameworks and hierarchical planning approaches support high-level reasoning, improve task decomposition, and reduce error propagation in complex workflows (2 studies, 1.75\%). Code inconsistency is mitigated through hybrid approaches that combine LLMs with engineered system components and traditional non-AI techniques, introducing dedicated agents to reason over code implementations and detect inconsistencies between intended behavior and actual code (2 studies, 1.75\%). To address the agent dependency challenge, agent monitoring mechanisms are employed as a practical solution (1 study, 0.88\%).
In addition, incorporating validation layers, fallback strategies, and interface checks is suggested to manage updates, deprecations, and breaking changes in APIs, libraries, and execution frameworks, thereby improving system reliability against tool changes and failures (1 study, 0.88\%). 
Finally, legacy code challenges are mitigated through intelligent maintenance solutions that assist in modernizing codebases and recommending updates to libraries and dependencies, ensuring alignment with contemporary software development standards (1 study, 0.88\%).

Overall, the reported challenges reveal that correctness and reliability issues in agent-based systems are not isolated model failures but systemic weaknesses emerging from the interaction between reasoning, coordination, and execution layers. The existing studies recommend that improving reliability in agent-based systems for code generation requires moving beyond scaling model capability and instead strengthening architectural governance, inter-agent accountability, and validation-aware design.

\begingroup
\renewcommand{\arraystretch}{1}
\begin{table*}[t]
\centering
\scriptsize
\caption{Identified challenges and their proposed solutions for LLM-based multi-agent systems for code generation}
\label{Tab: Challenges and Solutions}

\resizebox{\textwidth}{!}{%
\begin{tabular}{|l|l|l|l|l|}
\hline
\textbf{Category} & \textbf{Challenge Subcategory} & \textbf{Study ID (Challenge)} & \textbf{Solution} & \textbf{Study ID (Solution)} \\ \hline

% -------------------- Correctness and Reliability --------------------
\multirow{13}{*}{\textbf{\shortstack[l]{Correctness and Reliability\\Challenges with Agents}}}

& \multirow{3}{*}{Agent Orchestration Failure}
& \multirow{3}{*}{\shortstack[l]{\ref{P54}, \ref{P55}, \ref{P29}, \ref{G38}, \ref{G2}}}
& Schema-based communication protocols & \ref{P50}, \ref{G21} \\ \cline{4-5}

&  &  & Structured communication protocols & \ref{P50}, \ref{G21} \\ \cline{4-5}

&  &  & Dynamic communication structures & \ref{P50} \\ \cline{2-5}

& \multirow{3}{*}{Hallucination Challenge}
& \multirow{3}{*}{\shortstack[l]{\ref{P6}, \ref{P11}, \ref{P18}, \ref{P29},\\ \ref{P38}, \ref{P41}, \ref{P47}, \ref{P65},\\ \ref{G8}, \ref{G30}}}
& Integrating systematic error checking mechanism & \ref{P24} \\ \cline{4-5}

&  &  & Integrating feedback control mechanisms & \ref{P13} \\ \cline{4-5}

&  &  & Integrating automated verification tools & \ref{P13} \\ \cline{2-5}

& Code Inconsistency
& \shortstack[l]{\ref{P2}, \ref{P20}}
& Adopting hybrid approach & \ref{P56}, \ref{P20} \\ \cline{2-5}

& \multirow{2}{*}{Dependency Challenge}
& \multirow{2}{*}{\shortstack[l]{\ref{G8}, \ref{G10}, \ref{G30}}}
& Monitoring mechanisms & \ref{G10} \\ \cline{4-5}

&  &  & Validation layers & \ref{G30} \\ \cline{2-5}

& \multirow{2}{*}{Low Accuracy}
& \multirow{2}{*}{\shortstack[l]{\ref{P12}, \ref{P20}, \ref{P18}, \ref{P19},\\ \ref{P40}, \ref{G39}, \ref{G35}}}
& Integrating RL into agents & \ref{G35} \\ \cline{4-5}

&  &  & Iterative feedback driven mechanisms & \shortstack[l]{\ref{P57}, \ref{P19}, \ref{P40}} \\ \cline{2-5}

& Legacy Code
& \ref{G19}
& Code modernization & \ref{G19} \\ \cline{2-5}

& Lack of High-Level Planning
& \shortstack[l]{\ref{P1}, \ref{P65}, \ref{P72}, \ref{P48}}
& Proposed hierarchical planning approach & \ref{P48} \\ \hline

% -------------------- Security and Privacy --------------------
\multirow{7}{*}{\textbf{\shortstack[l]{Security and\\Privacy Risk}}}

& \multirow{2}{*}{Code Vulnerabilities}
& \multirow{2}{*}{\shortstack[l]{\ref{P31}, \ref{P58}, \ref{G32}}}
& Isolated environment & \shortstack[l]{\ref{P58}, \ref{P49}, \ref{G10}} \\ \cline{4-5}

&  &  & Control testing & \shortstack[l]{\ref{G10}, \ref{P58}} \\ \cline{2-5}

& \multirow{2}{*}{Code Attacks}
& \multirow{2}{*}{\shortstack[l]{\ref{G32}, \ref{G36}, \ref{P37}}}
& Monitoring mechanisms & \ref{G36} \\ \cline{4-5}

&  &  & External tool integration & \ref{G36} \\ \cline{2-5}

& \multirow{2}{*}{Data Leakage}
& \multirow{2}{*}{\shortstack[l]{\ref{G11}, \ref{G13}, \ref{G32}}}
& Open-source model control & \ref{G13} \\ \cline{4-5}

&  &  & Data loss prevention (DLP) mechanisms & \ref{G32} \\ \cline{2-5}

& Privacy
& \shortstack[l]{\ref{P58}, \ref{G2}, \ref{G9}, \ref{G10}}
& Security benchmarks & \ref{P58} \\ \hline

% -------------------- Computational and Resource Constraints --------------------
\multirow{4}{*}{\textbf{\shortstack[l]{Computational and\\Resource Constraints}}}
& Computational Cost
& \shortstack[l]{\ref{P8}, \ref{P12}, \ref{P18}, \ref{P38},\\ \ref{G2}, \ref{G5}, \ref{G15}, \ref{G39}, \ref{G27}}
& Utilizing small language models & \ref{G23} \\ \cline{2-5}

&  &  & Lightweight agents & \ref{P40} \\ \cline{2-5}

& Maintenance Overhead
& \shortstack[l]{\ref{G29}, \ref{G10}}
& Model simplification & \ref{G23} \\ \cline{2-5}

& Communication Overhead
& \shortstack[l]{\ref{G20}, \ref{G30}, \ref{G31}, \ref{P40}}
& Orchestration mechanisms & \shortstack[l]{\ref{P37}, \ref{G9}} \\ \hline

% -------------------- Context and Memory Limitations --------------------
\multirow{2}{*}{\textbf{\shortstack[l]{Context and Memory\\Limitations}}}
& Limited Context Window
& \shortstack[l]{\ref{G5}, \ref{G8}, \ref{G15}, \ref{P24},\\ \ref{P27}, \ref{P42}, \ref{P12}, \ref{P65}}
& External environment memory & \shortstack[l]{\ref{P23}, \ref{P59}} \\ \cline{2-5}

& Short-Term Memory
& \shortstack[l]{\ref{G6}, \ref{P37}, \ref{P64}}
& Explicit memory mechanisms & \shortstack[l]{\ref{G6}, \ref{P37}} \\ \hline

% -------------------- Benchmark Challenges --------------------
\multirow{3}{*}{\textbf{Benchmark Challenges}}
& Limited Real-World Evaluation
& \shortstack[l]{\ref{P17}, \ref{P62}, \ref{P70}, \ref{P12},\\ \ref{P53}, \ref{P23}}
& Repository Level Evaluation & \shortstack[l]{\ref{P17}, \ref{P70}} \\ \cline{2-5}

& Limited Language Scope
& \ref{P12}
& Cross-Lingual Benchmark & \ref{P17} \\ \cline{2-5}

& Security Risk
& \shortstack[l]{\ref{P12}, \ref{P68}}
& Secure Benchmark & \shortstack[l]{\ref{P58}, \ref{P17}} \\ \hline

% -------------------- Verification Challenges --------------------
\multirow{4}{*}{\textbf{Verification Challenges}}
& No Human Intervention
& \ref{P7}
& Hybrid verification approach & \ref{P65} \\ \cline{2-5}

& Non-Deterministic Behavior
& \shortstack[l]{\ref{G7}, \ref{G9}}
& Iterative verification cycle & \shortstack[l]{\ref{P57}, \ref{P65}} \\ \cline{2-5}

& Limited Self-Reflection
& \shortstack[l]{\ref{P54}, \ref{P72}}
& Symbolic reasoning & \shortstack[l]{\ref{P11}, \ref{P65}} \\ \cline{2-5}

& Lack of Task Validation
& \shortstack[l]{\ref{P6}, \ref{P65}, \ref{P69}, \ref{P72}, \ref{G21}}
& Integrating traditional tools into AI for validation & \ref{P65} \\ \hline

% -------------------- Prompt Design Challenges --------------------
\multirow{3}{*}{\textbf{Prompt Design Challenges}}
& Prompt Sensitivity
& \shortstack[l]{\ref{P29}, \ref{P6}, \ref{P47}, \ref{G8}}
& Agents’ multi-step reasoning & \ref{P42} \\ \cline{2-5}

& Prompt Dependency
& \ref{P19}
& Automatic prompt generation & \ref{P29} \\ \cline{2-5}

& Developer Prompt Interpretation Challenge
& \ref{P16}
& Human involvement in agent discussion & \ref{P29} \\ \hline

\end{tabular}%
}
\end{table*}
\endgroup

\subsubsection{Security and Privacy Risk}
%In this category, 
In this category, we report the security risks that threaten the confidentiality, integrity, and safe execution of code produced or executed by agent-based systems. In total, 11 out of 114 studies (9.65\%) report security-related challenges in agent-based code generation. The leading challenges are privacy risks (4 studies, 3.51\%), followed by code vulnerabilities (3 studies, 2.63\%), code attacks (3 studies, 2.63\%), and data leakage (3 studies, 2.63\%). To mitigate these challenges, the most commonly reported solutions include isolated execution environments (3 studies, 2.63\%), monitoring mechanisms across configuration files and generated code (1 studies, 0.88\%), and data protection measures such as DLP, audit logging, and secret/credential management (1 studies, 0.88\%). Other proposed solutions include deploying agents locally using open-source models to reduce data exposure (1 studies, 0.88\%), integrating external security tools to detect suspicious patterns, and using security-focused benchmarks and operational safeguards to evaluate and constrain risky agent behaviors in controlled settings (2 studies, 1.75\%). In summary, the evidence indicates that security risks are a recurring barrier to real-world deployment, and the reported solutions primarily emphasize containment, monitoring, and controlled evaluation rather than complete prevention.

\subsubsection{Computational and Resource Constraints}
This category reports resource allocation challenges that limit the ability of agent-based systems to efficiently manage computational resources during code generation tasks. A total of 15 out of 114 studies (13.16\%) highlight challenges related to high resource allocation in multi-agent systems. The leading challenges include computational cost (9 studies, 7.89\%), communication overhead (4 studies, 3.51\%), and maintenance overhead (2 studies, 1.75\%).
To mitigate these challenges, Small Language Models (SLMs) are suggested for agent development and execution in code and software development tasks to reduce computational costs (1 study, 0.88\%). Furthermore, optimized orchestration mechanisms are proposed to manage task dependencies, agent interactions, and monitoring and control processes in order to reduce communication overhead (2 studies, 1.75\%). In addition, a lightweight multi-agent framework is proposed for complex workflows or code generation, enabling targeted feedback and local test validation, thereby reducing token usage and avoiding unnecessary LLM calls during software development workflows (1 study, 0.88\%).

Overall, the reported challenges suggest that resource allocation limitations in agent-based systems are not only a matter of computational expense, but reflect architectural inefficiencies in coordination, communication, and task management. The findings highlight that sustainable agent-based code generation depends on balancing capability with computational efficiency through structured orchestration, modular design, and resource-aware execution mechanisms.

\subsubsection{Context and Memory Limitations}
In this category, we report memory-related challenges that limit the ability of LLM-based agent systems to maintain, retrieve, and utilize information across multi-step code generation workflows. In total, 11 out of 114 studies (9.65\%) report memory challenges in LLM-based agent systems for code generation and propose corresponding solutions. The reported challenges include limited context windows (8 studies, 7.02\%) and short-term memory limitations (3 studies, 2.63\%). 
To mitigate these challenges, the reported solutions primarily focus on extending and supporting agent memory. External memory environments are used to store and retrieve information beyond the model’s internal context (2 studies, 1.75\%). Additionally, explicit memory mechanisms are integrated to allow agents to store, reflect on, and reuse previously generated information during task execution (2 studies, 1.75\%). Overall, the findings indicate that memory limitations remain a fundamental bottleneck for agent-based systems in large and complex code generation tasks. The reported solutions suggest that augmenting agents with external and explicit memory mechanisms is important for supporting scalability and information reuse across multi-step workflows.

\subsubsection{Benchmark Challenges}
In this category, we report challenges related to the evaluation and benchmarking of agent-based systems for code generation. In this study, seven out of 114 studies (6.14\%) highlight challenges associated with existing benchmarks. These challenges include limited real-world evaluation benchmarks (6 studies, 5.26\%), limited language scope (1 study, 0.88\%), and benchmark security issues (2 studies, 1.75\%).
To mitigate challenges with existing benchmarks, SWE-bench \cite{jimenez2023swe}, LiveCodeBench \cite{jain2024livecodebench}, and CoderEval \cite{yu2024codereval} are proposed as steps toward repository-level, real-world, and context-aware evaluation frameworks, enabling evaluation in more realistic software development scenarios (2 studies, 1.75\%). In addition, benchmarks such as HumanEval-X \cite{peng2024humaneval}, MBXP \cite{athiwaratkun2022multi}, MultiPL-T \cite{cassano2022multipl}, and xCodeEval \cite{khan2023xcodeeval} extend traditional code generation tasks beyond a single programming language, enabling cross-lingual and multilingual evaluation of model performance (1 study, 0.88\%). To further address security-related challenges in existing benchmarks, REDCODE \cite{guo2024redcode} is proposed to assess the ability of models to recognize and handle unsafe code by providing high-quality, safety-oriented scenarios and test cases (2 studies, 1.75\%).

These findings indicate that existing benchmarks remain insufficient for comprehensively evaluating agent-based systems. The reported solutions highlight ongoing efforts toward more realistic, multilingual, and safety-aware benchmark designs to better reflect practical code generation scenarios. However, there remains a need for benchmarks and evaluation metrics that more comprehensively capture the complexity, variability, and dynamic nature of real-world software development workflows.

\subsubsection{Verification Challenges}
In this category, we report agent verification challenges, where limitations in verification mechanisms prevent the reliable validation of agent-generated code and agent interactions. 
In total, nine out of 114 studies (7.89\%) report challenges related to agent-based code verification and propose corresponding solutions. The reported challenges include non-deterministic or inconsistent agent behavior (2 studies, 1.75\%), lack of human intervention (1 study, 0.88\%), limited ability of agents to self-reflect (2 studies, 1.75\%), and lack of task validation (5 studies, 4.39\%). To address these challenges, symbolic reasoning tools are integrated into agents to validate logical assumptions in generated code (2 studies, 1.75\%).
In addition, a hybrid approach that combines agents with traditional software analysis techniques is proposed to automatically detect code defects and logic inconsistencies (1 study, 0.88\%).
Furthermore, a self-adaptive multi-agent framework is proposed to iteratively refine agent behaviors and interactions through testing and feedback-driven model updates, progressively improving code quality (1 study, 0.88\%).

The findings highlight that improving the agent verification process by implementing hybrid verification mechanisms, enhanced self-reflection, and feedback-driven adaptation is important to enhance agent reliability and code quality.

\subsubsection{Prompt Engineering Challenges}
Prompt engineering is an important aspect of LLM-based agents, as it directly influences how agents interpret instructions and generate code. In total, six out of 114 studies (5.26\%) report prompt engineering challenges, including prompt sensitivity (4 studies, 3.51\%), prompt dependency (1 study, 0.88\%), and developer prompt interpretation (1 study, 0.88\%). To mitigate these challenges, prompt strategies are proposed to enhance agents’ multi-step reasoning, enabling better interpretation of instructions and reducing errors in complex code generation tasks (1 study, 0.88\%). In addition, incorporating human involvement into agent discussions during the code generation process is suggested to improve understanding, guide decision-making, and reduce errors (1 study, 0.88\%).

In summary, the findings indicate that prompt engineering challenges affect agent reliability and correctness, particularly in complex code generation tasks. The proposed solutions take a step forward toward improving instruction interpretation and reducing errors, but remain limited in scope and generalizability.

\begin{tcolorbox}[width=\columnwidth, colback=gray!5, colframe=black!60, boxrule=0.5pt]
\textbf{Takeaway 5.}
The reviewed studies highlight seven major challenges in agent-based systems for code generation, which present significant barriers to real-world deployment. The reported solutions highlight a range of mitigation strategies, including real-time agent monitoring mechanisms, the use of sandboxed environments, adoption of local and small language models, integration of data protection measures such as DLP, optimized orchestration mechanisms, and techniques to improve accuracy and reliability. Together, these solutions reflect ongoing efforts to make agent-based code generation systems more practical and deployable in real-world settings.
\end{tcolorbox}

\subsection{Future Research Directions (RQ7)}
\label{Future Work}
This section presents the identified future work for LLM-based multi-agent systems for code generation, derived from the analysis of the selected studies. Based on the review, we identified six main categories of future research directions, which are summarized in Table \ref{Tab: future research}. More detailed descriptions can be found in the replication package \cite{rasheed_2026_18763362}.

\subsubsection{Advancing Agent Architecture and Optimization}
In this category, we report future directions focused on advancing agent system architecture and optimization to support more efficient, scalable, and robust agent-based code generation. In total, 15 out of 114 studies (13.16\%) highlight architectural and optimization-oriented directions aimed at improving the design, execution, and deployment of agent-based systems. These directions encompass improvements in agent scalability, orchestration mechanisms, edge deployment, parallel processing, fine-tuning and optimization strategies, runtime stability, and the integration of reinforcement learning to enhance coordination and decision-making.

Scalability is highlighted as a key future direction, with studies highlighting the need for scalable and token-efficient agent execution to support large and complex codebases. This points toward future work exploring approaches such as partial code representations, incremental updates, and selective context retrieval to reduce token consumption and computational overhead (4 studies, 3.51\%). In addition, adaptive scalability mechanisms that dynamically adjust the number of agents, memory usage, and coordination strategies based on task complexity and available resources are identified as important directions for further investigation (4 studies, 3.51\%). Agent orchestration is identified as an important future direction, focusing on structuring agent interactions through hierarchical coordination, summarization, and memory compression to enable efficient collaboration while limiting excessive context usage (4 studies, 3.51\%). In addition, future work indicates the need for improving agent communication through intelligent and adaptive orchestration mechanisms to reduce redundant interactions and unnecessary exchanges while preserving code generation quality, including adaptive communication control where agents selectively decide when and what to communicate based on task context, roles, and shared state (3 studies, 2.63\%).

Edge deployment is identified as an emerging future direction, with studies suggesting the exploration of lightweight and resource-efficient agent execution to support operation in constrained or low-connectivity environments, while potentially improving privacy, latency, and accessibility (1 study, 0.88\%). Parallel processing represents an important area for further investigation, as enabling the simultaneous execution of multiple models and tools within structured execution flows can improve efficiency, controllability, and reliability, while challenges related to coordination and synchronization remain to be addressed (1 study, 0.88\%). In addition, the fine-tuning and optimization of multi-agent systems require further research to support dynamic agent generation, workflow tuning, and iterative learning with human oversight and feedback, enabling a better balance between performance, computational cost, and latency (5 studies, 4.43\%). Runtime stability remains a critical issue, with future work needed to develop more reliable and sandbox-agnostic communication and execution monitoring mechanisms between agent controllers and execution environments (1 study, 0.88\%). Finally, reinforcement learning integration offers opportunities for improving coordination and decision-making in multi-agent systems by leveraging task-specific reward signals to guide execution, reduce error propagation, and improve logical consistency in generated programs (1 study, 0.88\%).

Overall, these directions highlight that advancing agent system architecture and optimization is essential for improving the scalability, efficiency, reliability, and practical deployability of agent-based code generation systems.

{\renewcommand{\arraystretch}{1}
\begin{table*}[t]
\centering
\scriptsize
\caption{Identified future research directions for LLM-based agent systems for code generation}
\label{Tab: future research}

\resizebox{\textwidth}{!}{%
\begin{tabular}{|l|l|l|l|}
\hline
\textbf{Category} & \textbf{Sub-Category} & \textbf{Details} & \textbf{Study ID} \\ \hline

\multirow{8}{*}{\begin{tabular}[c]{@{}l@{}}Advanced Agent Architecture \\ and  Optimization\end{tabular}}
& Scalability & Token-efficient execution and adaptive resource management & \ref{P10}, \ref{P24}, \ref{P52}, \ref{G28} \\ \cline{2-4}

& Agent orchestration & Intelligent and adaptive communication control & \ref{P7}, \ref{P20}, \ref{P50} \\ \cline{2-4}

& Edge deployment & Lightweight and resource-efficient agent execution & \ref{G13} \\ \cline{2-4}

& Parallel processing & Coordinated parallel execution of models and tools & \ref{P24} \\ \cline{2-4}

& Fine-tuning and optimization & Workflow tuning and iterative learning & \ref{P20}, \ref{P74}, \ref{P42}, \ref{P48}, \ref{P49} \\ \cline{2-4}

& Runtime stability & Sandbox-agnostic communication and execution monitoring & \ref{P49} \\ \cline{2-4}

& Reinforcement learning integration & Learning-based optimization and adaptive control & \ref{G11} \\ \cline{2-4}

& Context-aware agents & Context understanding and explainable reasoning & \ref{P24}, \ref{P4}, \ref{P74}, \ref{P2} \\ \hline

\multirow{4}{*}{Secure Agents}
& \multirow{2}{*}{Integrate privacy preserving}
& Integrate methods such as differential privacy, federated learning, and homomorphic encryption)
& \multirow{4}{*}{\begin{tabular}[c]{@{}l@{}}\ref{P52}, \ref{P53}, \ref{P63}, \ref{P66}, \ref{P74}, \\ \ref{P4}, \ref{P25}, \ref{P36}\end{tabular}} \\ \cline{3-3}

&  & Ensuring compliance with privacy regulations (e.g., GDPR, CCPA) &  \\ \cline{2-3}

& Permission allocation & Agent permission allocation and trust management &  \\ \cline{2-3}

& Enhanced Security & Expanding assessment criteria and use of isolated environment &  \\ \hline

\multirow{3}{*}{Smarter Memory}
& Multi-modal memory
& \begin{tabular}[c]{@{}l@{}}Multi-modal memory for cross-modal interaction in \\ LLM agents\end{tabular}
& \multirow{3}{*}{\ref{G14}, \ref{G13}, \ref{P24}, \ref{P52}} \\ \cline{2-3}

& Scalable memory management & Scalable memory through efficient LLM role management &  \\ \cline{2-3}

& Enhanced contextual models & Advancing memory-augmented for large-scale code &  \\ \hline

\multirow{3}{*}{Advanced Benchmarks}
& Multi-model benchmark & Multi-modal benchmark construction
& \multirow{3}{*}{\begin{tabular}[c]{@{}l@{}}\ref{P9}, \ref{P17}, \ref{P36}, \ref{P43}, \ref{P23}, \\ \ref{P67}, \ref{G28}, \ref{G16}\end{tabular}} \\ \cline{2-3}

& Secure benchmark & Safety, security, and reliability benchmarks &  \\ \cline{2-3}

& Agent communication & Agent communication and collaboration benchmarks &  \\ \hline

Multi-Model Integration
& Multi-modal agent architectures
& Expanding multi-modal LLM agents with human feedback
& \begin{tabular}[c]{@{}l@{}}\ref{P5}, \ref{P24}, \ref{P49}, \ref{P12}, \ref{P36}, \\ \ref{G11}, \ref{G13}\end{tabular} \\ \hline

Human--AI Collaboration
& Human in loop with agents
& Human-in-the-loop feedback and decision support
& \ref{P1}, \ref{P2}, \ref{P12}, \ref{P30}, \ref{G10}, \ref{G28} \\ \hline

\end{tabular}
}
\end{table*}}

\subsubsection{Secure Agents for Future Development}
In total, eight out of 114 studies (7.02\%) highlight future research directions aimed at enhancing the security of agents. Enhancing the security of agents requires specialized agents to handle security-critical and non-functional requirements, enabling context-aware delegation of security-sensitive tasks within sandboxed environments without disrupting developer workflows (5 studies, 4.39\%).
In addition, privacy-preserving collaboration is identified as an important direction, calling for the integration of techniques such as differential privacy and secure computation to protect sensitive data during agent interactions (1 study, 0.88\%).
Permission allocation and trust management are identified as key directions, suggesting controlled collaboration mechanisms such as role-based access and trust-aware coordination to reduce security risks in multi-agent systems (2 studies, 1.75\%).

In summary, the findings indicate that advancing agent security requires systematic integration of specialized security roles, controlled access mechanisms, and privacy-aware architectures to support safe and trustworthy deployment in real-world SE contexts.

\subsubsection{Smarter Memory Mechanisms}
In total, four out of 114 studies (3.51\%) highlight smarter memory mechanisms as a key future direction for agent-based systems. These research directions include multi-modal and structured memory systems that store multi-modal data alongside summaries, entities, relationships, and procedural knowledge derived from documents, code, and interactions (2 studies, 1.75\%). In addition, scalable memory management is identified as an important research direction, suggesting the exploration of adaptive mechanisms that dynamically expand or compress memory and computational resources based on task demands and agent activity (2 studies, 1.75\%). Furthermore, enhanced contextual models that combine larger context windows with memory-augmented architectures are highlighted to support reasoning over large and complex codebases, enabling the capture of long-range dependencies and hierarchical reasoning across multiple levels of abstraction (1 study, 0.88\%).
Overall, the findings highlight that advancing agent memory systems can be achieved by integrating multi-modal and structured memory representations with scalable, adaptive management mechanisms, enabling more effective reasoning over complex, multi-modal information and large codebases.

\subsubsection{Advanced Benchmarks}
\label{Advance Benchmark}
% \textbf{Realistic and open-ended design benchmarks}:
% Current benchmark such as Swe-bench, LiveCodeBench and 

%  Future benchmarks should therefore move toward more realistic and complex design problems that incorporate long-term dependencies, system-level constraints, and practical engineering considerations encountered in real-world development environments. 

% \textbf{Standardized and comprehensive evaluation frameworks.}
% There is a need for standardized benchmark frameworks that systematically evaluate LLMs and agents across diverse SE activities, particularly in requirements engineering. Future benchmarks should assess agents’ abilities to interpret stakeholder needs, resolve conflicting requirements, and generate consistent specifications. In addition, evaluating non-functional requirements, such as performance, security, and reliability, remains an underexplored yet critical direction with strong implications for industrial adoption.

% \textbf{Agent-centric performance and robustness evaluation.}
% Future benchmark research should focus on evaluating agent performance across dimensions such as consistency, reliability, adaptability, and reusability across tasks and domains. Benchmarks should support comparative analyses of different agent architectures, collaboration protocols, and decision-making strategies, while also enabling the exploration of alternative execution paths and outcomes. Such evaluations are essential for understanding the strengths and limitations of LLM-based agents beyond task-level accuracy.
In total, eight out of 114 studies (7.02\%) highlight future research required for advanced benchmarks. These include multi-modal benchmark construction, where benchmarks incorporate structured multi-modal inputs to provide a more in-depth assessment of cross-modal reasoning and generation abilities (2 studies, 1.75\%). In addition, future benchmarks should focus on real-world tasks and incorporate evaluations of safety, security, and reliability (2 studies, 1.75\%). Furthermore, advanced benchmarks should assess collaborative capabilities among multiple agents, including coordination efficiency, communication effectiveness, task decomposition quality, reduction in human effort, as well as logical correctness, control flow, and structural reasoning (4 studies, 3.51\%).
These findings highlight the need for more realistic and multi-dimensional benchmark designs. Such benchmarks are essential for systematically evaluating agent performance beyond isolated tasks and for advancing robust, collaborative, and reliable agent-based systems.

\subsubsection{Multi-Model Agent Systems}
%this section we can take help from the paper "From LLMs to LLM-based Agents for Software Engineering: A Survey of Current, Challenges and Future" This paper wrote future work that can help us to write this section.

%As shown in Table X, seven studies highlight the importance of multi-model integration in software development. 
In this study, seven out of 114 studies (6.14\%) report that future research should focus on multi-modal agent architectures, where agents can process and understand text, images, audio, and video by integrating vision–language models with other modality-specific components, such as browser-based visual perception and code-driven processing of structured files (e.g., XLSX). Furthermore, expanding multi-modal LLM-based agents by incorporating human feedback mechanisms represents a promising direction for advancing engineering design automation and complex software development tasks.
The interest in multi-modal agent architectures suggests an emerging shift from text-only code generation toward more context-aware and environment-integrated systems.

\subsubsection{Human-AI Collaboration}

In this study, six out of 114 studies (5.26\%) report that a key future research direction is to strengthen human–AI collaboration in LLM-based agent systems by integrating human feedback and external knowledge into the decision-making process, enabling agents to handle routine analysis and implementation while humans provide guidance, validation, and high-level decision-making. Future systems are therefore likely to adopt hybrid models in which agents perform routine analysis and implementation tasks, while humans provide oversight, contextual understanding, and strategic decision-making, indicating a shift toward collaborative rather than purely autonomous development workflows.

\begin{tcolorbox}[width=\columnwidth, colback=gray!5, colframe=black!60, boxrule=0.5pt]
\textbf{Takeaway 6.}
This study categorizes future research on LLM-based agent systems for code generation into six key areas, highlighting major challenges and opportunities for real-world deployment. These comprise improving reliability, security, memory mechanisms, benchmarking, human–AI collaboration, and multi-model systems. Addressing these is essential to move agent-based code generation from experimental prototypes to practical, production-ready SE tools.
\end{tcolorbox}
\section{Results Discussion and Analysis}
\label{Discussions}
In this section, we first discuss the results, which are organized around the research questions to interpret the key findings and identify existing gaps as well as implications for future research (Section \ref{Results Discussion}). Then, in Section \ref{CurrentStateofResearchandPractice}, we provide an analysis of the current state of research and practice on LLM-based multi-agent systems for code generation.

\subsection{Results Discussion}
\label{Results Discussion}

\subsubsection{Demography (RQ1)} 
To understand both academic and industry perspectives on LLM-based multi-agent systems for code generation, we included both peer-reviewed and grey literature. As shown in Figure \ref{fig:demography_white}, 42 studies were published by academia in 2024, while only six studies were published by industry. This indicates that LLM-based multi-agent systems were explored more actively by academia than by industry during that year. In contrast, we observed a substantial increase in industry contributions in 2025, with 34 studies published by industry compared to 21 studies from academia. This pattern suggests that the topic was initially driven by academic research, with industry engagement gaining momentum at a later stage. This may indicate that the challenges associated with multi-agent systems, such as security, reliability, and trustworthiness, have limited their adoption in real-world industrial settings or resulted in comparatively lower initial interest from industry. In addition, the analysis shows that industry publications more frequently highlight security-related aspects, whereas academic studies more commonly focus on correctness and reliability in agent-based systems. This difference in focus further highlights the priorities of the two sectors, with industry concentrating more on deployment-related risks, while academia focuses on technical validation and performance evaluation.
Furthermore, this study highlights the limited collaboration between academia and industry. Only 7.89\% of the studies involve collaboration between academic and industrial institutions, whereas 57.02\% of the studies are affiliated with academia and 35.09\% with industry. This distribution shows that, although both academia and industry are actively contributing to this area, collaboration between them remains limited. 

\textbf{Implications}:For researchers, the findings highlight that strengthening collaboration between academia and industry would improve the realism of experimental designs and improve the practical relevance of research outcomes and deployment contexts. In addition, for practitioners, the growth of peer-reviewed research provides a structured and increasingly well-organized body of knowledge, offering clearer insights into publication trends, research focus areas, and the ongoing development of LLM-based multi-agent systems for code generation across both academic and industrial domains.

\subsubsection{Reasons to use Agents (RQ2)} 
For this RQ, we analyzed the studies to understand the reasons for using multi-agent systems for code generation. As shown in Table \ref{tab:reason}, we categorized the identified reasons into the following nine categories: performance enhancement, complex task handling, operational efficiency and practical scalability, team collaboration, context management, external tool integration, autonomous execution, adaptability, and future-oriented development. Overall, the results suggest that organizations and researchers are leveraging agent architectures to address the limitations observed in single-model approaches, particularly in managing complexity, scalability, and real-world deployment demands. As mentioned in studies such as \cite{he2025llm}, \cite{zhang2024codeagent}, \cite{shu2024towards}, and \cite{talebirad2023multi}, collaborative and tool-augmented agent systems improve performance, contextual awareness, and execution reliability compared to standalone LLMs. These studies also highlight that the growing interest in autonomous agents for the software development process represents a step forward toward more structured and production-ready AI-assisted programming solutions.

\textbf{Implications}: For practitioners, the findings suggest that adopting agent-based systems can provide strategic advantages in managing complex development tasks, improving productivity, and enhancing system performance. However, successful implementation requires careful architectural design, workflow integration, and control mechanisms to ensure reliability and cost-effectiveness in real-world environments. For researchers, the results highlight that multi-agent systems improve performance compared to single-agent models; however, further investigation is required to understand how agent architectures can systematically enhance coordination, adaptability, and scalability in SE tasks. The challenges of agent-based systems, as highlighted in Section \ref{chalengess and Solutions}, should be further investigated and explored to improve their reliability and practical applicability.

\subsubsection{Benchmarks and Models (RQ3-RQ4)}
% Overall findings,
% The limtations
% And then implications

%Same for models as well. 
%I want the Discussion to be written so that: It discusses the key takeaways for each research question. For each research question, the discussion should include at least: a) Interpretation of the key results (what the results mean), b) Connections to existing literature (showing whether the findings confirm, extend, or contradict prior work), c) Implications for both researchers and practitioners, d) Future research directions, if applicable. Your task is to: List the key findings for each research question, and Write the discussion for one key finding using the structure above.

\textbf{Benchmarks} play an important role in evaluating the code generation capabilities of LLM-based agent systems. In this study, we identified 37 benchmarks that have been utilized to assess LLM-based agents in software development tasks. Among these, HumanEval \cite{chen2021evaluating} is the most frequently used benchmark by both academia and industry, while MBPP \cite{austin2021program} is the second most commonly adopted benchmark. These results align with prior studies, which highlight HumanEval and MBPP as two of the earliest standardized benchmarks for code generation, contributing to their widespread adoption due to their simplicity, well-defined evaluation protocols, and unit-test–based assessment of functional correctness \cite{yu2024humaneval}, \cite{paul2024benchmarks}.
However, recently developed benchmarks such as SWE-bench and LiveCodeBench aim to evaluate agent performance in more realistic, repository-level, and context-rich settings, but their adoption remains limited compared to traditional unit-test–based benchmarks. Detailed information about the identified benchmarks, including their evaluation focus and usage context, is provided in Table \ref{tab:merged_benchmark_table_final}.

\textbf{Implications}: For practitioners, the findings suggest that reported benchmark performance may not directly translate to real-world software development scenarios, where factors such as code quality, maintainability, security, and system-level behavior are critical. As a result, practitioners should interpret benchmark results with caution when considering the adoption of agent-based code generation systems and carefully assess whether the evaluated benchmarks reflect their intended deployment context.

For researchers, our analysis highlights several benchmark-related challenges that require further investigation. Existing benchmarks often have limited real-world representativeness, lack scalability to large or long-horizon tasks, involve high computational costs, and provide insufficient evaluation of multi-agent collaboration and coordination. These limitations reduce the ability of current benchmarks to measure the capabilities of LLM-based agent systems accurately. To address these issues, we discussed advanced benchmarking directions in Section~\ref{Advance Benchmark}, where we propose that future research should focus on developing realistic, open-ended, and task-diverse benchmarks that better reflect industrial SE practices. In addition, there is a need for multi-modal benchmark designs that incorporate code, documentation, issue-tracking data, and execution environments. Researchers should also prioritize agent-centric evaluation criteria, such as collaboration efficiency, behavior under failure, adaptability to changing requirements, and long-term task completion, to support more systematic and meaningful evaluation of LLM-based agent systems.

\textbf{Models} are a foundational component of LLM-based agent systems, as their capabilities directly influence agent performance, reliability, and applicability in software development tasks. As illustrated in Table \ref{tab:llm_frequency_table10}, most reviewed studies rely on closed-source models, with OpenAI’s GPT series being the most frequently adopted. Open-source models, including Llama, Mixtral, and DeepSeek, are also employed; however, their adoption remains comparatively limited.  
These results align with prior work showing that closed-source models, such as the GPT family, often outperform open-source alternatives on benchmarks and integration tasks \cite{laskar2023building}, \cite{kulkarni2025blue}.
However, as reported in the above findings, the use of closed-source models introduces challenges related to security risks, potential data leakage, and increased operational costs (see Section~\ref{chalengess and Solutions}). In contrast, open-source models reduce security risks by enabling deployment in local, isolated environments, while also improving data control, enhancing transparency, and lowering overall costs.

\textbf{Implications}: For practitioners, the findings suggest that, although closed-source models offer strong performance, high efficiency, and ease of integration, their adoption in real-world systems requires careful consideration of security risks, data privacy, and cost constraints. In contrast, open-source models are recommended for practical deployment scenarios where security, data control, and privacy are critical, particularly when models can be deployed in local or isolated environments. For researchers, these findings highlight that future research should systematically assess open-source models in realistic and secure deployment settings, while also exploring techniques to narrow the performance gap and preserve data control and reproducibility.

\subsubsection{Challenges and Solutions (RQ5-RQ6)}
While LLM-based multi-agent systems demonstrate strong potential to accelerate code generation, improve software development efficiency, and support complex tasks such as multi-file and modular software architectures, they also face significant challenges. These challenges include security and privacy risks, agent reliability and correctness issues, high computational and resource costs, limitations in long-term memory handling, verification difficulties, and the lack of realistic real-world evaluation benchmarks. The reviewed studies propose solutions to mitigate these limitations, as summarized in Section \ref{chalengess and Solutions}.
These challenges are key empirical findings of this study, as they highlight critical gaps in the practical deployment of LLM-based agents. As mentioned in the literature \cite{sypherd2024practical, yehudai2025survey, liang2025llm, he2025llm, wang2025comprehensive}, current agent systems are primarily adopted for prototype-level development; however, their applicability in real-world deployment scenarios remains limited. This indicates that, despite promising experimental results, existing agent-based systems currently lack the level of reliability and scalability required for use in production environments. Overall, these findings suggest that advancing LLM-based agents beyond experimental settings requires greater attention to system-level design, evaluation, and validation. 
At a systems level, the concentration of challenges around reliability, security, orchestration, memory management, and evaluation suggests that many current limitations arise less from model capability and more from gaps in engineering infrastructure and architectural standardization. This indicates that the field is shifting from a model-centric phase, focused mainly on improving reasoning performance, to a systems engineering phase, where coordination stability, verification pipelines, cost governance, and secure deployment become key challenges. Consequently, the progress of multi-agent systems will depend less on stronger underlying models and more on the development of well-defined design patterns, standardized orchestration strategies, and reproducible evaluation frameworks that support production-level deployment.

\textbf{Implications}: For practitioners, the findings of this study suggest that the practical deployment of agent-based systems requires careful implementation through appropriate design choices that balance security, cost, and reliability, rather than relying solely on model capability. In industrial settings, the adoption of LLM-based agents should be accompanied by proper security controls, including isolated sandboxed execution environments, real-time monitoring mechanisms, data loss prevention (DLP) techniques to mitigate sensitive data leakage, and systematic review of AI-generated code. From a system design perspective, practitioners should incorporate explicit memory management strategies, such as external memory stores or retrieval-based mechanisms, to support long-horizon tasks and maintain contextual consistency, as well as effective agent orchestration solutions, including role-based coordination, task decomposition, and controlled inter-agent communication, to reduce error propagation and improve overall system reliability. To address cost-related constraints, this study recommends the use of smaller language models and local deployments where feasible, alongside hybrid human–agent workflows to reduce coordination complexity and ongoing maintenance overhead. Finally, to improve agent accuracy and reliability in real-world environments, practitioners are encouraged to integrate continuous monitoring, automated validation, and human-in-the-loop review mechanisms throughout the software development lifecycle.

For researchers, the findings of this study highlight the need for further empirical investigations into the security and reliability of LLM-based agent systems, particularly to assess attack surfaces, vulnerability propagation, and the effectiveness of mitigation strategies in realistic software development scenarios. Future research should also focus on improving agent scalability by reducing communication overhead, enhancing orchestration strategies, and systematically balancing performance gains against computational and resource costs. In addition, empirical studies that evaluate resource–performance trade-offs in real-world industrial settings are necessary to support the development of more efficient and sustainable agent-based systems. Finally, to strengthen agent verification and evaluation, researchers are encouraged to develop advanced benchmarking frameworks that assess agent performance under realistic, long-horizon, and multi-agent conditions reflective of real-world SE workflows.

\subsubsection{Future Trends}
% Here we say that above we found that many of direction for future trend and then we explain below one by one. 
This study provides evidence-based insights into the real-world applicability of agent-based code generation systems and highlights several promising directions for future development, as discussed in detail in Section~\ref{Future Work}. We have summarized the key emerging trends that are likely to shape the next generation of LLM-based agent systems. These include the integration of multiple models within agent architectures, enhanced security and privacy considerations for agents, the development of smarter and scalable memory systems, the development of more advanced, realistic, and multi-modal benchmarking approaches, improved mechanisms for agent monitoring and orchestration, stronger human–AI collaboration frameworks, and the adoption of more effective and cost-efficient fine-tuning strategies. 
These directions indicate that the field is moving beyond isolated architectural experimentation toward the development of more integrated agent ecosystems. While continued improvements in model capability remain relevant, future progress will depend more strongly on how effectively model performance is combined with orchestration logic, memory infrastructure, security controls, and evaluation pipelines within unified system designs. This suggests that the next stage of development requires stronger architectural alignment and clearer standardization, particularly in benchmarking practices, coordination mechanisms, and deployment governance.

\textbf{Implications}:
For practitioners, these directions provide guidance for the responsible and effective adoption of agent-based code generation systems in real-world settings. They highlight the importance of moving beyond experimental prototypes toward scalable and production-ready solutions, while maintaining appropriate oversight and governance mechanisms. 
For researchers, these directions outline a forward-looking research agenda that encourages the systematic advancement of agent architectures and evaluation methodologies, ensuring that future developments are more powerful, reliable, trustworthy, and deployable in practical settings.

\subsection{{Current State of Research and Practice}}
\label{CurrentStateofResearchandPractice}

This subsection synthesizes evidence from both academic literature and grey literature to characterize the current state of research and practice in LLM-based multi-agent system for code generation. We move beyond performance comparisons and assess the systemic maturity of the research ecosystem. Using the evidence synthesized across RQ2--RQ7, we examine whether LLM-based multi-agent shows the characteristics of a mature ecosystem or remains in an earlier developmental phase.

\textbf{Industry--academia alignment}: The reviewed studies indicate that multi-agent systems are proposed to address complex programming and SE tasks. As synthesized in RQ2, both academic and industrial sources highlight the handling of complex tasks, team collaboration, and reasoning depth as key motivations for adopting multi-agent architectures.
However, regarding RQ3, we found that 29 studies used HumanEval and 18 studies used MBPP as primary evaluation benchmarks. These benchmarks are designed for function-level code generation and relatively small programming tasks. They involve short problem descriptions, isolated functions, and limited contextual reasoning. This creates a structural misalignment between research ambition and evaluation practice. While multi-agent systems are positioned as solutions for complex SE challenges, their performance is largely assessed on simplified benchmark tasks.

% \textcolor{blue}{
% More specifically, we observe an \emph{ambition--evaluation misalignment}: while multi-agent systems are framed as solutions for complex, collaborative, and long-horizon SE tasks (RQ2), their empirical validation predominantly relies on short, function-level benchmarks such as HumanEval and MBPP (RQ3). These benchmarks do not evaluate sustained coordination, architectural reasoning, repository-scale modification, or integration with development workflows.
% }

In addition, a \emph{research--deployment misalignment} is evident. Academic evaluations primarily focus on functional correctness, whereas industrial adoption requires security validation, compliance assurance, and cost control. These structural tensions indicate that evaluation practices have not yet evolved to match real-world deployment conditions. These gaps suggest that the research ecosystem has not yet fully aligned with industrial realities.

% Furthermore, organizations must consider cost efficiency, scalability, and compliance with internal security policies when deploying automated coding agents. Security and cost related challenges are mostly highlighted by industry in the grey literature.

\textbf{Reproducibility and model dependence}:
Reproducibility is another important indicator of systemic maturity. Our results from RQ3 show that 32 studies utilized earlier model versions, such as GPT-3.5 or GPT-4. As a result, experiments conducted with these earlier model configurations cannot be exactly reproduced under current conditions. This creates a temporal reproducibility challenge, where reported results become difficult to verify or replicate as the underlying language models evolve.
Additionally, the dependence on closed-source models limits experimental transparency and comparability across studies. Since these models are controlled by external providers, researchers and practitioners cannot fully access model parameters or guarantee stable behavior over time. A mature research ecosystem typically supports stable baselines, controlled evaluation environments, and the ability to reproduce experiments independently. When results depend on evolving external APIs, long-term replication becomes challenging for both academic researchers and industrial teams attempting to validate or adopt proposed systems. In industrial settings, reproducibility is essential for internal validation, auditing, and compliance processes. The absence of standardized reporting of prompts, model versions, and experimental configurations further weakens reproducibility and makes cross-study comparison difficult. These conditions suggest that the field remains in a stage of rapid experimentation rather than methodological stabilization.

This pattern reflects an \emph{innovation--standardization tension}, where rapid model evolution and experimentation outpace the establishment of stable baselines, standardized reporting practices, and long-term reproducibility norms.

\textbf{Pathways toward systemic maturity}:
The combined evidence from benchmark usage, models reproducibility practices, industry misalignment, and the existing challenges highlighted in Section \ref{chalengess and Solutions} indicates that multi-agent systems are currently in an advanced prototyping or early consolidation phase.
This finding aligns with previous studies \cite{he2025llm, han2024llm, elhashemy2025bridging, liu2024large}, which describe multi-agent systems as research-driven prototypes that have not yet reached production-ready deployment maturity.

From a systemic perspective, this phase is characterized by high architectural heterogeneity, absence of agreed benchmark standards beyond function-level tasks, strong dependence on evolving proprietary models, and limited deployment-oriented validation. These characteristics are typical of a pre-standardization ecosystem transitioning from exploratory prototyping toward early consolidation, rather than a fully mature and stabilized research domain.

For the ecosystem to progress toward systemic maturity, several structural developments are necessary. First, evaluation practices must better reflect real-world SE complexity.
In addition, security assessment should become an integral component of evaluation, including the incorporation of safe execution constraints. Controlled and secure agent environments are particularly important when autonomous agents interact with external tools or repositories.

Furthermore, moving toward a more mature multi-agent ecosystem requires operational transparency and deployment awareness. Clear documentation of model versions, prompts, architectural configurations, and system-level coordination mechanisms is essential to support consistent validation across different environments. In multi-agent settings, coordination strategies often increase computational overhead, making scalability and resource management central concerns for practical adoption. A mature system should therefore incorporate systematic reporting of computational demands, latency characteristics, and deployment constraints, ensuring that performance claims extend beyond accuracy metrics to reflect real-world implementation conditions.
Finally, the development of multi-model and heterogeneous agent systems may contribute to greater maturity. Many current systems rely on a single underlying language model with role-based prompting. Future systems may incorporate specialized models for planning, coding, security validation, and testing within controlled environments. Such structured integration could improve reliability and align more closely with real-world SE practices.

Overall the evidence across benchmarks (RQ3), model usage patterns (RQ4), identified challenges (Section~\ref{chalengess and Solutions}), and future research directions (RQ7) suggests that the field exhibits strong innovation momentum but limited structural consolidation. Architectural experimentation is widespread, yet standardized coordination patterns have not stabilized. Evaluation practices remain narrow relative to system ambitions. Reproducibility is limited by frequent model changes and incomplete reporting. These signals collectively indicate that LLM-based multi-agent system has not yet reached full ecosystem-level maturity, but rather represents an evolving domain undergoing structural negotiation between research exploration and emerging deployment demands.

By addressing these structural challenges, the field can transition from innovation-driven experimentation toward a more stable, reproducible, production-ready, and deployment-aware research ecosystem.
\section{Threats to Validity}
\label{Threats to Validity}
We followed the guidelines proposed by Runeson \textit{et al.} \cite{runeson2009guidelines} to identify and discuss potential threats to the validity of the study. This section is structured into four categories: construct validity, internal validity, external validity, and conclusion validity.

\subsection{Internal validity}
According to the Runeson \textit{et al}. \cite{runeson2009guidelines}, internal validity concerns factors that may affect the correctness of data extraction and analysis performed on the selected studies. In this MLR, threats to internal validity may arise at different stages of the research process.

\begin{itemize}
    \item \textbf{Study Search}: There is a risk of missing relevant and important academic and industrial studies during the search process, which requires careful attention. To reduce this risk, we applied an integrated use of primary search and snowballing techniques, as described in Section \ref{Search Strategy}. To maximize the retrieval of primary studies during the initial search phase, we refined the search string through pilot searches and multiple iterations in collaboration with the second and third authors. The finalized search string was then applied to the selected databases, enabling an effective and well-defined search across both academic and industrial sources.

    \item \textbf{Study Selection}: As this study includes both peer-reviewed and grey literature, the screening and selection process was particularly critical. To address this concern, we defined a structured study screening and selection procedure, which is summarized in Table 3. To reduce personal bias, we first conducted an initial screening of the studies, followed by a qualitative assessment of the selected papers. Throughout this process, the first two authors performed the screening in accordance with the criteria explicitly specified in Table 3.

    \item \textbf{Data Extraction}: In an MLR, data extraction can be challenging due to the possibility of researcher bias. To mitigate this concern, we designed a standardized data extraction form (see Table \ref{Tab Dataitem}) to support consistent data collection. The initial data extraction was carried out by the first author, after which the second and third authors reviewed the results. Any discrepancies were discussed and resolved collaboratively.

    \item \textbf{Data Analysis}: In this study, both qualitative and quantitative methods were applied to analyze the collected data. Biases introduced during data analysis may affect the interpretation of the results. To mitigate this risk, we employed an open coding technique derived from Blair \textit{et al}. \cite{blair2015reflexive}. In addition, the data analysis was reviewed by the second and third authors, and any disagreements were discussed and resolved.

    \item \textbf{Domain-Specific Threats to Internal Validity.}
    This domain introduces additional technical threats to internal validity. Many primary studies rely on API-based proprietary LLMs whose architectures, training data, and internal configurations are not publicly disclosed. The black-box nature of these systems limits transparency in experimental settings and may complicate accurate interpretation of reported results. Furthermore, variations in evaluation protocols, such as differences in prompt design, temperature settings, pass@k calculation methods, and tool integration strategies, may introduce inconsistencies across studies. These methodological variations may affect the comparability and interpretation of performance outcomes synthesized in this review.
    
\end{itemize}

\subsection{External validity}
External validity concerns the extent to which the study results can be generalized \cite{runeson2009guidelines}. To strengthen external validity, we established a detailed study protocol that clearly defines the research methodology. The study includes both peer-reviewed and grey literature published between January 2022 and June 2025, collected from five academic databases and three additional sources. These sources cover research on LLM-based agents for code generation and are summarized in Table \ref{tab:sources}. 

As shown in Table \ref{tab:llm_frequency_table10}, many studies evaluated earlier GPT models (e.g., GPT-3.5 and Codex/Davinci). Given the rapid evolution of LLMs, where models are continuously updated, fine-tuned, or replaced, the reported performance and behavioral characteristics may not generalize to newer versions. Consequently, findings derived from earlier model generations may not fully transfer to more recent or substantially modified LLMs. Furthermore, several industrial studies rely on proprietary datasets, internal development environments, or organization-specific workflows, which may limit the generalizability of reported productivity gains, reliability improvements, or scalability claims to other contexts, programming languages, or organizational settings.

\subsection{Construct validity}
Construct validity refers to the extent to which a measurement or instrument accurately represents and captures the theoretical concept or construct it is intended to measure \cite{wohlin2014guidelines}. In this study, construct validity primarily concerns the potential subjectivity involved in the analysis of the selected studies. To mitigate this risk, data extraction was conducted independently by the first author, after which the second and third authors reviewed the extracted data. Any disagreements were resolved through discussion, following the recommendations of Kitchenham’s guidelines \cite{kitchenham2009systematic}. In addition, the quality of each selected study was evaluated in accordance with the assessment protocol proposed by Dybå and Dingsøyr \cite{dybaa2008empirical}.

\subsection{Conclusion validity}
Conclusion validity refers to the reliability of the conclusions derived from the results \cite{wohlin2014guidelines}. In this study, potential threats to conclusion validity include the possible omission of relevant studies. To reduce this risk, we adhered to established best practices, such as defining a rigorous search protocol and conducting pilot searches and pilot data selection, as recommended by Keele \textit{et al}. \cite{keele2007guidelines}. Furthermore, to strengthen the reliability of the conclusions, the authors collaborated closely to interpret the findings and agree on the final conclusions jointly.
\section{Related Work}
\label{background}

In this section, we discuss the literature relevant to our study. Section~\ref{LLMsinSE} summarizes studies investigating the use of LLMs in SE, whereas Section~\ref{LLMs-based agents in SE} examines research on LLM-based agent systems within the SE domain. 

\subsection{Large Language Models for Software Engineering}
\label{LLMsinSE} 

Several surveys and literature reviews have examined the integration and application of LLMs in SE, synthesizing current advancements, identifying open challenges, and outlining emerging research trends \cite{quin2024b, zheng2023survey, ozkaya2023application, fan2023large, hou2023large, zhang2023critical, husein2025large, wang2024software, zheng2025towards, shi2025efficient, wang2024software, gormez2024large, albuquerque2024generating, zhang2024systematic, DBLP:journals/corr/abs-2509-13144}. For example, Zheng \textit{et al}. \cite{zheng2023survey} presented one of the earliest survey papers in 2023, which reviewed and assessed 134 studies focusing on code-related LLMs. Their analysis examined the relationship between general purpose and code specialized models and evaluated their performance across a wide range of SE tasks, providing detailed insights into the capabilities and limitations of LLMs in this domain. In the same year, Fan \textit{et al}. \cite{fan2023large} presented a survey that summarized existing research and highlighted open research challenges related to the application of LLMs to technical problems faced in SE. Additionally, Hou \textit{et al}. \cite{hou2023large} conducted a systematic literature review on LLMs for SE, in which they analyzed 395 research articles and categorized the different LLMs that have been employed across various SE tasks.

In 2024, the volume of research on LLMs for SE increased substantially, indicating a growing interest in both general and task-specific applications of LLMs. Building on an earlier survey, Hou \textit{et al}. \cite{quin2024b} conducted a large-scale review of studies published between 2017 and 2023, examining how different types of LLMs have been applied across a wide range of SE tasks. Their analysis covered data processing pipelines, optimization strategies, and application domains, while also identifying key challenges, open research gaps, and future research directions.
At a broader level, Gormez \textit{et al}. \cite{gormez2024large} presented a systematic mapping study that characterized the application landscape of LLMs in SE, with particular attention to their capabilities and limitations, including issues related to non-determinism and hallucinations. Similarly, Albuquerque \textit{et al}. \cite{albuquerque2024generating} conducted a systematic mapping study of 19 selected studies, identifying 23 LLMs, 13 utilization strategies, 15 challenges, and 14 mitigation mechanisms associated with the use of LLMs in programming workflows, thereby providing a structured overview of current practices.

As the field matured, several studies in 2024 shifted the focus toward specific SE sub-domains, particularly software testing and automated program repair. Wang \textit{et al}. \cite{wang2024software} reviewed 102 studies that investigated the use of LLMs in software testing, finding that tasks such as test case generation and program repair are among the most common applications. Focusing specifically on code repair, Zhang \textit{et al}. \cite{zhang2024systematic} presented the first systematic literature review dedicated to the application of LLMs in automated program repair, synthesizing evidence from 189 relevant papers and highlighting the remaining challenges as well as directions for future research.

In early 2025, Zheng \textit{et al}. \cite{zheng2025towards} presented a comprehensive review of 123 studies to examine the current integration of LLMs within SE and to assess whether LLMs effectively support SE tasks. Their work provided an overview of how LLMs are incorporated into SE workflows by organizing SE tasks into multiple categories and discussing representative application scenarios. The study highlighted both the strengths and limitations of existing approaches and identified persistent challenges, including performance variability, insufficient evaluation of code-oriented models, and the need for task-specific model customization.
More recently, Husein \textit{et al}. \cite{husein2025large} conducted a systematic literature review on the use of LLMs for code completion, a fundamental SE task aimed at improving developer productivity through context-aware code prediction. Their analysis of 23 primary studies demonstrated that LLMs can significantly enhance code completion performance, while also revealing open challenges and outlining directions for future research.

\subsection{Large Language Model-Based Agents in Software Engineering}
\label{LLMs-based agents in SE}
LLM-based multi-agent systems interact and coordinate to support collaborative decision-making in complex tasks \cite{DBLP:conf/ijcai/GuoCWCPCW024}. In this domain, several literature reviews and surveys have been conducted to synthesize existing research on LLM-based agents and their role in SE. For instance, in 2024, Jin \textit{et al}. \cite{jin2024llms} examined the evolution of agent-based systems by identifying key components such as planning, memory, tool usage, and feedback mechanisms. In the same year, Li \textit{et al}. \cite{li2024survey} surveyed LLM-based multi-agent systems and proposed a unified workflow-oriented structure composed of five core components, while also summarizing their applications in problem-solving and world simulation and discussing the key challenges in this area. Together, these studies provided a structured view of agent architectures and coordination mechanisms in LLM-based systems.
Building on these efforts, Wang \textit{et al}. \cite{wang2025agents} extended the discussion in 2025 by offering a broader overview of agent paradigms across the SE lifecycle, situating LLM-based agents within established research traditions and outlining emerging opportunities and challenges. Additionally, Xi \textit{et al}. \cite{xi2025rise} presented a survey of LLM-based agents, covering architectural frameworks, application scenarios, and the emergence of agent societies, while also identifying open research problems and future research directions.

As discussed above, research activity continued to grow throughout late 2024 and into 2025, leading to the publication of more comprehensive and specialized surveys, including those by He \textit{et al}. \cite{he2025llm} and Liu \textit{et al}. \cite{liu2024large}. These studies synthesized an expanding body of academic studies on role-based collaboration, agent orchestration strategies, evaluation methodologies, and scalability challenges. In parallel, more focused reviews, such as that by Dong \textit{et al}. \cite{dong2025survey}, examined the application of agentic systems to programming tasks, comparing single-agent and multi-agent approaches for code synthesis, debugging, and automated program repair. 

\subsection{Conclusive Summary}
\label{Conclusive Summary}
Existing reviews primarily synthesize peer-reviewed academic literature and typically adopt a broad SE or general LLM perspective. Most surveys treat code generation as one of many application areas and do not examine it in depth as a primary task within multi-agent system architectures. Moreover, there is a clear lack of studies that systematically incorporate grey literature to explore industrial perspectives on agent-based code generation. As a result, practitioner-driven innovations, industrial tool designs, implementation strategies, and real-world deployment experiences receive limited attention in prior reviews. This gap restricts the field’s understanding of how multi-agent architectures for code generation are actually engineered, evaluated, and adopted in practice. By explicitly focusing on LLM-based multi-agent systems for code generation and systematically integrating both academic and grey literature, this MLR provides a broader and more practice-oriented synthesis of practical challenges, emerging patterns, and adoption trends, thereby advancing the current understanding of the field.

\section{Conclusions}
\label{Conclusions}
This MLR presents the current state of research on LLM–based multi-agent systems for code generation from both academic and industrial perspectives. The primary objective of this study was to examine the motivations for employing multi-agent systems in code generation, identify commonly used benchmarks for evaluating agent performance, and analyze the models adopted by academia and industry to execute multi-agent systems. In addition, this MLR synthesizes the key challenges associated with multi-agent systems, along with their proposed solutions in the literature. We also highlight emerging trends and future research directions for further investigation.
To address the research questions outlined in Table~\ref{RQs}, we analyzed a total of 114 studies, including 74 peer-reviewed publications and 40 sources from the grey literature. The key findings of this MLR include:

\begin{itemize}
    \item The findings of this MLR present several insights into the demographics of the reviewed peer-reviewed and grey literature, including publishers, publication types, and authors’ and organizational affiliations. The yearly distribution of peer-reviewed publications reached its peak in 2024, with 42 papers, while the grey literature peaked in 2025 with 33 sources, indicating a growing and continuing research interest in LLM-based multi-agent systems. In peer-reviewed studies, publications are largely dominated by Google Scholar–indexed venues (39.2\%), with conferences serving as the primary avenue for publications. Similarly, in the grey literature, Google-based sources also dominate, representing approximately 65\% of the collected studies.

    \item This MLR identifies several motivations for adopting LLM-based multi-agent systems in code generation and categorizes them into nine groups: operational efficiency and practical scalability, handling complex tasks, performance enhancement, context management, external tool integration, team collaboration, autonomous execution, adaptability, and pathways toward AGI and emerging future trends. These findings are important for both researchers and practitioners, as they clarify the key drivers behind the adoption of multi-agent approaches and inform the design of more effective and scalable agentic systems.

   \item We have identified 37 benchmarks that are used to evaluate the performance of LLM-based agents for code generation. Among these, HumanEval is the most frequently adopted benchmark, used 29 times to assess agent-based system performance, followed by MBPP, which was used 18 times. Table~\ref{tab:merged_benchmark_table_final} provides a detailed overview of benchmark usage across the reviewed studies. We have also highlighted the open challenges associated with existing benchmarks, including limited coverage of real-world scenarios and insufficient evaluation of security and privacy aspects.
    In addition, we analyzed the models employed across the selected studies to better understand current execution practices. We found that closed-source models were used 57 times, whereas open-source models were adopted 26 times in the selected studies, indicating a strong reliance on proprietary models for agent execution. Among these, the OpenAI GPT series emerged as the most widely used family of models, appearing 43 times across the reviewed literature, reflecting its central role in current agent-based code generation research.

    \item In this study, we have identified and categorized seven key challenges associated with LLM-based multi-agent systems for code generation. These challenges include correctness and reliability challenges with agents, security and privacy risks, computational and resource constraints, context and memory limitations, benchmark challenges, verification, and prompt design challenges. We have also summarized the mitigation strategies and proposed solutions discussed in the literature (see Section~\ref{chalengess and Solutions}). These findings highlight the need for more reliable agent architectures, improved evaluation methodologies, and stronger safeguards to support the reliable and secure deployment of agentic systems in real-world SE scenarios.

    \item In this MLR, we identify future research directions, which are categorized into six areas aimed at further investigating and improving LLM-based multi-agent systems for code generation. These directions include advanced agent architectures, secure and reliable agents, the development of smarter memory mechanisms for agents, advanced benchmarks, multi-model integration within agentic frameworks, and human–AI collaboration. Together, these directions provide a roadmap for advancing the efficiency and practical applicability of agentic systems in software development.

\end{itemize}

The findings of this MLR will benefit researchers by providing an overview of the current state of research on LLM-based multi-agent systems for code generation and by identifying open research challenges for further investigation, as discussed in Section~\ref{Future Work}. In addition, the insights identified from this study support knowledge transfer to practitioners by providing a clear overview of existing challenges, proposed solutions, and emerging trends in LLM-based agentic frameworks for code generation. We highlight the importance of practitioners considering these findings when designing, deploying, and maintaining agent-based systems, particularly with respect to security, scalability, reliability, and cost in real-world software development contexts.

\appendix
% \appendices
% See Table \ref{tab:Abbreviations}
\label{sec:Appendix}

\section*{Acknowledgment}
This project was co-funded by the MAISA project (2025–2027), funded by Business Finland, which represents a collaboration between academia and eight leading Finnish companies.

\section*{Data availability}

To facilitate replication and validation of this study, we have publicly released the dataset ~\cite{rasheed_2026_18763362}.

\section*{Declaration of AI Assistance}

During the preparation of this manuscript, the author(s) used ChatGPT to assist with grammar refinement, sentence restructuring, and formatting improvements. Following the use of this tool, the author(s) carefully reviewed and revised the content and assume full responsibility for the final version of the publication.
% \bibliographystyle{unsrtnat}
% \bibliography{references}  %%% Uncomment this line and comment out the ``thebibliography'' section below to use the external .bib file (using bibtex) .

% \bibliographystyle{unsrtnat}
% \bibliography{original_studies}
\section*{Selected Studies}

\begin{enumerate}[label=P\arabic*]
\item \label{P1}
Huan Zhang, Wei Cheng, Yuhan Wu, and Wei Hu.
\textit{A pair programming framework for code generation via multi-plan exploration and feedback-driven refinement}.
In Proceedings of the 39th IEEE/ACM International Conference on Automated Software Engineering,
pages 1319--1331, 2024.

\item \label{P2}
Yuntong Zhang, Haifeng Ruan, Zhiyu Fan, and Abhik Roychoudhury.
\textit{Autocoderover: Autonomous program improvement}.
In Proceedings of the 33rd ACM SIGSOFT International Symposium on Software Testing and Analysis,
pages 1592--1604, 2024.

\item \label{P3}
Shubham Gandhi, Manasi Patwardhan, Lovekesh Vig, and Gautam Shroff.
\textit{BudgetMLAgent: A cost-effective LLM multi-agent system for automating machine learning tasks}.
In Proceedings of the 4th International Conference on AI-ML Systems,
pages 1--9, 2024.

\item \label{P4}
Peya Mowar, Yi-Hao Peng, Jason Wu, Aaron Steinfeld, and Jeffrey P. Bigham.
\textit{CodeA11y: Making AI coding assistants useful for accessible web development}.
In Proceedings of the 2025 CHI Conference on Human Factors in Computing Systems,
pages 1--15, 2025.

\item \label{P5}
Chen-Chia Chang, Chia-Tung Ho, Yaguang Li, Yiran Chen, and Haoxing Ren.
\textit{DRC-Coder: Automated DRC checker code generation using LLM autonomous agent}.
In Proceedings of the 2025 International Symposium on Physical Design,
pages 143--151, 2025.

\item \label{P6}
Anastasiia Grishina, Vadim Liventsev, Aki Härmä, and Leon Moonen.
\textit{Fully autonomous programming using iterative multi-agent debugging with large language models}.
ACM Transactions on Evolutionary Learning, 5(1):1--37, 2025.

\item \label{P7}
Jie Wu and Fatemeh H. Fard.
\textit{HumanEvalComm: Benchmarking the communication competence of code generation for LLMs and LLM agents}.
ACM Transactions on Software Engineering and Methodology, 34(7):1--42, 2025.

\item \label{P8}
Hao Ding, Ziwei Fan, Ingo Guehring, Gaurav Gupta, Wooseok Ha, Jun Huan,
Linbo Liu, Behrooz Omidvar-Tehrani, Shiqi Wang, and Hao Zhou.
\textit{Reasoning and planning with large language models in code development}.
In Proceedings of the 30th ACM SIGKDD Conference on Knowledge Discovery and Data Mining,
pages 6480--6490, 2024.

\item \label{P9}
Nathaniel Pinckney, Christopher Batten, Mingjie Liu, Haoxing Ren, and Brucek Khailany.
\textit{Revisiting VerilogEval: A year of improvements in large-language models for hardware code generation}.
ACM Transactions on Design Automation of Electronic Systems, 2025.

\item \label{P10}
Yihong Dong, Xue Jiang, Zhi Jin, and Ge Li.
\textit{Self-collaboration code generation via ChatGPT}.
ACM Transactions on Software Engineering and Methodology, 33(7):1--38, 2024.

\item \label{P11}
Mark Marron.
\textit{Toward programming languages for reasoning: Humans, symbolic systems, and AI agents}.
In Proceedings of the 2023 ACM SIGPLAN International Symposium on New Ideas, New Paradigms,
and Reflections on Programming and Software,
pages 136--152, 2023.

\item \label{P12}
Meghana Puvvadi, Sai Kumar Arava, Adarsh Santoria,
Sesha Sai Prasanna Chennupati, and Harsha Vardhan Puvvadi.
\textit{Coding agents: A comprehensive survey of automated bug fixing systems and benchmarks}.
In 2025 IEEE 14th International Conference on Communication Systems and Network Technologies (CSNT),
pages 680--686, 2025.

\item \label{P13}
Jingzhi Gong, Vardan Voskanyan, Paul Brookes, Fan Wu, Wei Jie, Jie Xu,
Rafail Giavrimis, Mike Basios, Leslie Kanthan, and Zheng Wang.
\textit{Language models for code optimization: Survey, challenges and future directions}.
arXiv preprint arXiv:2501.01277, 2025.

\item \label{P14}
Levent Dinçkal.
\textit{Large language model-based autonomous agents: Trends and directions}.
AIPA's International Journal on Artificial Intelligence: Bridging Technology, Society and Policy,
1(1):13--24, 2024.

\item \label{P15}
Junda He, Christoph Treude, and David Lo.
\textit{LLM-Based Multi-Agent Systems for Software Engineering: Literature Review, Vision, and the Road Ahead}.
ACM Transactions on Software Engineering and Methodology, 34(5):1--30, 2025.

\item \label{P16}
Peter Robe, Sandeep K. Kuttal, Jake AuBuchon, and Jacob Hart.
\textit{Pair programming conversations with agents vs.\ developers: challenges and opportunities for SE community}.
In Proceedings of the 30th ACM Joint European Software Engineering Conference and Symposium on the Foundations of Software Engineering,
pages 319--331, 2022.

\item \label{P17}
Kaixin Wang, Tianlin Li, Xiaoyu Zhang, Chong Wang, Weisong Sun, Yang Liu, and Bin Shi.
\textit{Software Development Life Cycle Perspective: A Survey of Benchmarks for Code Large Language Models and Agents}.
arXiv preprint arXiv:2505.05283, 2025.

\item \label{P18}
Rolando Ram{\'\i}rez-Rueda, Edgard Ben{\'\i}tez-Guerrero, Carmen Mezura-Godoy, and Everardo B{\'a}rcenas.
\textit{Transforming Software Development: A Study on the Integration of Multi-Agent Systems and Large Language Models for Automatic Code Generation}.
In 2024 12th International Conference in Software Engineering Research and Innovation (CONISOFT),
pages 11--20, 2024.

\item \label{P19}
Haolin Jin, Zechao Sun, and Huaming Chen.
\textit{RGD: Multi-LLM based agent debugger via refinement and generation guidance}.
In 2024 IEEE International Conference on Agents (ICA),
pages 136--141, 2024.

\item \label{P20}
Xingyuan Bai, Shaobin Huang, Chi Wei, and Rui Wang.
\textit{Collaboration between intelligent agents and large language models: A novel approach for enhancing code generation capability}.
Expert Systems with Applications, 269:126357, 2025.

\item \label{P21}
Rafael Barbarroxa, Luis Gomes, and Zita Vale.
\textit{Benchmarking large language models for multi-agent systems: A comparative analysis of AutoGen, CrewAI, and TaskWeaver}.
In International Conference on Practical Applications of Agents and Multi-Agent Systems,
pages 39--48, 2024.

\item \label{P22}
Buvaneswari A. Ramanan, Manzoor A. Khan, and Abhinav Rao.
\textit{ASPIRE: A Multi-Agent Framework for Execution-Free Code Analysis and Repair}.
In 2024 IEEE International Conference on Big Data (BigData),
pages 8811--8813, 2024.

\item \label{P23}
Kechi Zhang, Jia Li, Ge Li, Xianjie Shi, and Zhi Jin.
\textit{CodeAgent: Enhancing Code Generation with Tool-Integrated Agent Systems for Real-World Repo-level Coding Challenges}.
In Proceedings of the 62nd Annual Meeting of the Association for Computational Linguistics (ACL 2024),
pages 13643--13658, 2024.

\item \label{P24}
Samuel Holt, Max Ruiz Luyten, and Mihaela van der Schaar.
\textit{L2MAC: Large Language Model Automatic Computer for Extensive Code Generation}.
In The Twelfth International Conference on Learning Representations (ICLR 2024),
2024.

\item \label{P25}
Md.\ Ashraful Islam, Mohammed Eunus Ali, and Md.\ Rizwan Parvez.
\textit{MapCoder: Multi-Agent Code Generation for Competitive Problem Solving}.
In Proceedings of the 62nd Annual Meeting of the Association for Computational Linguistics (ACL 2024),
pages 4912--4944, 2024.

\item \label{P26}
S.\ Akilesh, Rajeev Sekar, Om Kumar C.\ U., Prakalya D., and Suguna M.
\textit{Multi-Agent hierarchical workflow for autonomous code generation with Large Language Models}.
In 2025 IEEE International Students' Conference on Electrical, Electronics and Computer Science (SCEECS),
pages 1--5, 2025.

\item \label{P27}
Yusen Zhang, Ruoxi Sun, Yanfei Chen, Tomas Pfister, Rui Zhang, and Sercan Arik.
\textit{Chain of agents: Large language models collaborating on long-context tasks}.
Advances in Neural Information Processing Systems, 37:132208--132237, 2024.

\item \label{P28}
Zhuofan Shi, Chunxiao Xin, Tong Huo, Yuntao Jiang, Bowen Wu, Xingyue Chen, Wei Qin, Xinjian Ma, Gang Huang, Zhenyu Wang, and others.
\textit{A fine-tuned large language model based molecular dynamics agent for code generation to obtain material thermodynamic parameters}.
Scientific Reports, 15(1):10295, 2025.

\item \label{P29}
Lei Wang, Chen Ma, Xueyang Feng, Zeyu Zhang, Hao Yang, Jingsen Zhang, Zhiyuan Chen, Jiakai Tang, Xu Chen, Yankai Lin, and others.
\textit{A survey on large language model based autonomous agents}.
Frontiers of Computer Science, 18(6):186345, 2024.

\item \label{P30}
Zeeshan Rasheed, Muhammad Waseem, Malik Abdul Sami, Kai-Kristian Kemell, Aakash Ahmad, Anh Nguyen Duc, Kari Syst{\"a}, and Pekka Abrahamsson.
\textit{Autonomous agents in software development: A vision paper}.
In International Conference on Agile Software Development,
pages 15--23, 2024.

\item \label{P31}
Thiem Nguyen Ba, Binh Nguyen Thanh, and Viet-Trung Tran.
\textit{CoverNexus: Multi-agent LLM System for Automated Code Coverage Enhancement}.
In International Symposium on Information and Communication Technology,
pages 472--484, 2024.

\item \label{P32}
Dawei Yuan, Guocang Yang, and Tao Zhang.
\textit{UI2HTML: Utilizing LLM agents with chain of thought to convert UI into HTML code}.
Automated Software Engineering, 32(2):1--24, 2025.

\item \label{P33}
Kaiyan Chang, Wenlong Zhu, Kun Wang, Xinyang He, Nan Yang, Zhirong Chen, Dantong Jin, Cangyuan Li, Yunhao Zhou, Yan Hao, and others.
\textit{A data-centric chip design agent framework for Verilog code generation}.
ACM Transactions on Design Automation of Electronic Systems, 2025.

\item \label{P34}
Haoyuan Wu, Zhuolun He, Xinyun Zhang, Xufeng Yao, Su Zheng, Haisheng Zheng, and Bei Yu.
\textit{ChatEDA: A Large Language Model Powered Autonomous Agent for EDA}.
IEEE Transactions on Computer-Aided Design of Integrated Circuits and Systems, 43(10):3184--3197, 2024.

\item \label{P35}
Chen Qian, Wei Liu, Hongzhang Liu, Nuo Chen, Yufan Dang, Jiahao Li, Cheng Yang, Weize Chen, Yusheng Su, Cong Xin, and others.
\textit{ChatDev: Communicative agents for software development}.
In Proceedings of the 62nd Annual Meeting of the Association for Computational Linguistics (Volume 1: Long Papers),
pages 15174--15186, 2024.

\item \label{P36}
Hanbin Wang, Zhenghao Liu, Shuo Wang, Ganqu Cui, Ning Ding, Zhiyuan Liu, and Ge Yu.
\textit{Intervenor: Prompting the coding ability of large language models with the interactive chain of repair}.
In Findings of the Association for Computational Linguistics: ACL 2024,
pages 2081--2107, 2024.

\item \label{P37}
Robert Feldt, Sungmin Kang, Juyeon Yoon, and Shin Yoo.
\textit{Towards autonomous testing agents via conversational large language models}.
In 2023 IEEE/ACM International Conference on Automated Software Engineering (ASE),
pages 1688--1693, 2023.

\item \label{P38}
Taicheng Guo, Xiuying Chen, Yaqi Wang, Ruidi Chang, Shichao Pei, Nitesh V.\ Chawla, Olaf Wiest, and Xiangliang Zhang.
\textit{Large Language Model Based Multi-agents: A Survey of Progress and Challenges}.
In Proceedings of the Thirty-Third International Joint Conference on Artificial Intelligence (IJCAI 2024),
pages 8048--8057, 2024.

\item \label{P39}
Carlos E.\ Jimenez, John Yang, Alexander Wettig, Shunyu Yao, Kexin Pei, Ofir Press, and Karthik R.\ Narasimhan.
\textit{SWE-bench: Can Language Models Resolve Real-world GitHub Issues?}
In The Twelfth International Conference on Learning Representations (ICLR 2024),
2024.

\item \label{P40}
Dong Huang, Jie M.\ Zhang, Michael Luck, Qingwen Bu, Yuhao Qing, and Heming Cui.
\textit{AgentCoder: Multi-agent-based code generation with iterative testing and optimisation}.
arXiv preprint arXiv:2312.13010, 2023.

\item \label{P41}
Qian Huang, Jian Vora, Percy Liang, and Jure Leskovec.
\textit{Benchmarking large language models as AI research agents}.
In NeurIPS 2023 Foundation Models for Decision Making Workshop,
2023.

\item \label{P42}
Xingyao Wang, Yangyi Chen, Lifan Yuan, Yizhe Zhang, Yunzhu Li, Hao Peng, and Heng Ji.
\textit{Executable code actions elicit better LLM agents}.
In Proceedings of the Forty-first International Conference on Machine Learning (ICML),
2024.

\item \label{P43}
Qingyun Wu, Gagan Bansal, Jieyu Zhang, Yiran Wu, Beibin Li, Erkang Zhu, Li Jiang,
Xiaoyun Zhang, Shaokun Zhang, Jiale Liu, and others.
\textit{AutoGen: Enabling next-gen LLM applications via multi-agent conversations}.
In Proceedings of the First Conference on Language Modeling,
2024.

\item \label{P44}
Chia-Tung Ho, Haoxing Ren, and Brucek Khailany.
\textit{VerilogCoder: Autonomous Verilog coding agents with graph-based planning and abstract syntax tree (AST)-based waveform tracing tool}.
In Proceedings of the AAAI Conference on Artificial Intelligence,
39(1):300--307, 2025.

\item \label{P45}
John Yang, Carlos E. Jimenez, Alexander Wettig, Kilian Lieret, Shunyu Yao,
Karthik Narasimhan, and Ofir Press.
\textit{SWE-Agent: Agent-computer interfaces enable automated software engineering}.
Advances in Neural Information Processing Systems, 37:50528--50652, 2024.

\item \label{P46}
Islem Bouzenia, Premkumar T. Devanbu, and Michael Pradel.
\textit{RepairAgent: An Autonomous, LLM-Based Agent for Program Repair}.
In 47th IEEE/ACM International Conference on Software Engineering (ICSE 2025),
pages 2188--2200, 2025.

\item \label{P47}
Victor Dibia, Jingya Chen, Gagan Bansal, Suff Syed, Adam Fourney, Erkang Zhu,
Chi Wang, and Saleema Amershi.
\textit{AUTOGEN STUDIO: A No-Code Developer Tool for Building and Debugging Multi-Agent Systems}.
In Proceedings of the 2024 Conference on Empirical Methods in Natural Language Processing: System Demonstrations,
pages 72--79, 2024.

\item \label{P48}
Guohao Li, Hasan Hammoud, Hani Itani, Dmitrii Khizbullin, and Bernard Ghanem.
\textit{CAMEL: Communicative agents for ``mind'' exploration of large language model society}.
Advances in Neural Information Processing Systems, 36:51991--52008, 2023.

\item \label{P49}
Xingyao Wang, Boxuan Li, Yufan Song, Frank F. Xu, Xiangru Tang, Mingchen Zhuge,
Jiayi Pan, Yueqi Song, Bowen Li, Jaskirat Singh, Hoang H. Tran, Fuqiang Li,
Ren Ma, Mingzhang Zheng, Bill Qian, Yanjun Shao, Niklas Muennighoff, Yizhe Zhang,
Binyuan Hui, Junyang Lin, Robert Brennan, Hao Peng, Heng Ji, and Graham Neubig.
\textit{OpenHands: An Open Platform for AI Software Developers as Generalist Agents}.
In The Thirteenth International Conference on Learning Representations (ICLR 2025),
2025.

\item \label{P50}
Weize Chen, Yusheng Su, Jingwei Zuo, Cheng Yang, Chenfei Yuan, Chi-Min Chan,
Heyang Yu, Yaxi Lu, Yi-Hsin Hung, Chen Qian, and others.
\textit{AgentVerse: Facilitating Multi-Agent Collaboration and Exploring Emergent Behaviors}.
In The Twelfth International Conference on Learning Representations (ICLR),
2024.

\item \label{P51}
Yujie Zhao, Hejia Zhang, Hanxian Huang, Zhongming Yu, and Jishen Zhao.
\textit{Mage: A multi-agent engine for automated RTL code generation}.
In 2025 62nd ACM/IEEE Design Automation Conference (DAC),
pages 1--7, 2025.

\item \label{P52}
Yuheng Cheng, Ceyao Zhang, Zhengwen Zhang, Xiangrui Meng, Sirui Hong, Wenhao Li,
Zihao Wang, Zekai Wang, Feng Yin, Junhua Zhao, and others.
\textit{Exploring large language model based intelligent agents: Definitions, methods, and prospects}.
arXiv preprint arXiv:2401.03428, 2024.

\item \label{P53}
Junda He, Christoph Treude, and David Lo.
\textit{LLM-Based Multi-Agent Systems for Software Engineering: Literature Review, Vision, and the Road Ahead}.
ACM Transactions on Software Engineering and Methodology, 34(5):1--30, 2025.

\item \label{P54}
Sirui Hong, Mingchen Zhuge, Jonathan Chen, Xiawu Zheng, Yuheng Cheng, Jinlin Wang,
Ceyao Zhang, Zili Wang, Steven Ka Shing Yau, Zijuan Lin, and others.
\textit{MetaGPT: Meta programming for a multi-agent collaborative framework}.
In The Twelfth International Conference on Learning Representations (ICLR),
2023.

\item \label{P55}
Dong Chen, Shaoxin Lin, Muhan Zeng, Daoguang Zan, Jian-Gang Wang, Anton Cheshkov,
Jun Sun, Hao Yu, Guoliang Dong, Artem Aliev, and others.
\textit{Coder: Issue resolving with multi-agent and task graphs}.
arXiv preprint arXiv:2406.01304, 2024.

\item \label{P56}
Gang Fan, Xiaoheng Xie, Xunjin Zheng, Yinan Liang, and Peng Di.
\textit{Static code analysis in the AI era: An in-depth exploration of the concept, function, and potential of intelligent code analysis agents}.
arXiv preprint arXiv:2310.08837, 2023.

\item \label{P57}
Yue Hu, Yuzhu Cai, Yaxin Du, Xinyu Zhu, Xiangrui Liu, Zijie Yu, Yuchen Hou,
Shuo Tang, and Siheng Chen.
\textit{Self-Evolving Multi-Agent Collaboration Networks for Software Development}.
In The Thirteenth International Conference on Learning Representations (ICLR 2025),
2025.

\item \label{P58}
Chengquan Guo, Xun Liu, Chulin Xie, Andy Zhou, Yi Zeng, Zinan Lin, Dawn Song, and Bo Li.
\textit{RedCode: Risky Code Execution and Generation Benchmark for Code Agents}.
In Advances in Neural Information Processing Systems (NeurIPS 2024),
2024.

\item \label{P59}
Haolin Jin, Linghan Huang, Haipeng Cai, Jun Yan, Bo Li, and Huaming Chen.
\textit{From LLMs to LLM-based Agents for Software Engineering: A Survey of Current, Challenges and Future}.
arXiv preprint arXiv:2408.02479, 2024.

\item \label{P60}
Xiangru Tang, Yuliang Liu, Zefan Cai, Yanjun Shao, Junjie Lu, Yichi Zhang, Zexuan Deng,
Helan Hu, Kaikai An, Ruijun Huang, Shuzheng Si, Chen Sheng, Haozhe Zhao, Liang Chen,
Tianyu Liu, Yin Fang, Yujia Qin, Wangchunshu Zhou, Yilun Zhao, Zhiwei Jiang,
Baobao Chang, Arman Cohan, and Mark Gerstein.
\textit{ML-Bench: Evaluating Large Language Models for Code Generation in Repository-Level Machine Learning Tasks}.
OpenReview preprint, 2025.

\item \label{P61}
Anonymous.
\textit{SWE-PolyBench: A multi-language benchmark for repository level evaluation of coding agents}.
Submitted to The Fourteenth International Conference on Learning Representations (under review),
2025.

\item \label{P62}
Kaiyuan Liu, Youcheng Pan, Yang Xiang, Daojing He, Jing Li, Yexing Du, and Tianrun Gao.
\textit{ProjectEval: A Benchmark for Programming Agents Automated Evaluation on Project-Level Code Generation}.
In Findings of the Association for Computational Linguistics: ACL 2025,
pages 20205--20221, 2025.

\item \label{P63}
Niels M{\"u}ndler, Mark Niklas M{\"u}ller, Jingxuan He, and Martin T. Vechev.
\textit{SWT-Bench: Testing and Validating Real-World Bug-Fixes with Code Agents}.
In Advances in Neural Information Processing Systems (NeurIPS 2024),
2024.

\item \label{P64}
Chen Qian, Yufan Dang, Jiahao Li, Wei Liu, Zihao Xie, Yifei Wang, Weize Chen,
Cheng Yang, Xin Cong, Xiaoyin Che, Zhiyuan Liu, and Maosong Sun.
\textit{Experiential Co-Learning of Software-Developing Agents}.
In Proceedings of the 62nd Annual Meeting of the Association for Computational Linguistics (ACL 2024),
pages 5628--5640, 2024.

\item \label{P65}
Samdyuti Suri, Sankar Narayan Das, Kapil Singi, Kuntal Dey, Vibhu Saujanya Sharma, and Vikrant Kaulgud.
\textit{Software Engineering Using Autonomous Agents: Are We There Yet?}
In 38th IEEE/ACM International Conference on Automated Software Engineering (ASE 2023),
pages 1855--1857, 2023.

\item \label{P66}
Dong Huang, Qingwen Bu, Jie M. Zhang, Michael Luck, and Heming Cui.
\textit{AgentCoder: Multi-Agent-based Code Generation with Iterative Testing and Optimisation}.
arXiv preprint arXiv:2312.13010, 2023.

\item \label{P67}
Xunzhu Tang, Kisub Kim, Yewei Song, Cedric Lothritz, Bei Li, Saad Ezzini, Haoye Tian,
Jacques Klein, and Tegawend{\'e} F. Bissyand{\'e}.
\textit{CodeAgent: Autonomous Communicative Agents for Code Review}.
In Proceedings of the 2024 Conference on Empirical Methods in Natural Language Processing (EMNLP 2024),
pages 11279--11313, 2024.

\item \label{P68}
Godsfavour Kio.
\textit{SWE-bench-secret: Automating AI Agent Evaluation for Software Engineering Tasks}.
University of Waterloo, 2025.

\item \label{P69}
Niels M{\"u}ndler, Mark Niklas M{\"u}ller, Jingxuan He, and Martin T. Vechev.
\textit{Code agents are state of the art software testers}.
In ICML 2024 Workshop on LLMs and Cognition,
2024.

\item \label{P70}
Jirat Pasuksmit, Wannita Takerngsaksiri, Patanamon Thongtanunam, Chakkrit Tantithamthavorn,
Ruixiong Zhang, Shiyan Wang, Fan Jiang, Jing Li, Evan Cook, Kun Chen, and others.
\textit{Human-In-The-Loop Software Development Agents: Challenges and Future Directions}.
In 2025 IEEE/ACM 22nd International Conference on Mining Software Repositories (MSR),
pages 756--757, 2025.

\item \label{P71}
Junwei Liu, Kaixin Wang, Yixuan Chen, Xin Peng, Zhenpeng Chen, Lingming Zhang, and Yiling Lou.
\textit{Large language model-based agents for software engineering: A survey}.
arXiv preprint arXiv:2409.02977, 2024.

\item \label{P72}
Chunqiu Steven Xia, Yinlin Deng, Soren Dunn, and Lingming Zhang.
\textit{Demystifying LLM-based software engineering agents}.
Proceedings of the ACM on Software Engineering, 2(FSE):801--824, 2025.

\item \label{P73}
Md.\ Ashraful Islam, Mohammed Eunus Ali, and Md.\ Rizwan Parvez.
\textit{CodeSim: Multi-Agent Code Generation and Problem Solving through Simulation-Driven Planning and Debugging}.
In Findings of the Association for Computational Linguistics: NAACL 2025,
pages 5113--5139, 2025.

\item \label{P74}
Yiming Huang, Jianwen Luo, Yan Yu, Yitong Zhang, Fangyu Lei, Yifan Wei, Shizhu He,
Lifu Huang, Xiao Liu, Jun Zhao, and Kang Liu.
\textit{DA-Code: Agent Data Science Code Generation Benchmark for Large Language Models}.
In Proceedings of the 2024 Conference on Empirical Methods in Natural Language Processing (EMNLP 2024),
pages 13487--13521, 2024.

\end{enumerate}

\begin{enumerate}[label=G\arabic*]

\item \label{G1}
Yoichi Ishibashi and Yoshimasa Nishimura.
\textit{Self-organized agents: A LLM multi-agent framework toward ultra large-scale code generation and optimization}.
arXiv preprint arXiv:2404.02183, 2024.

\item \label{G2}
GetStream.
\textit{Best 5 Frameworks to Build Multi-Agent AI Applications}.
Online article, 2025.
Available at: \url{https://getstream.io/blog/multiagent-ai-frameworks/}.

\item \label{G3}
Jagadeesan Ganesh.
\textit{Mastering LLM AI Agents: Building and Using AI Agents in Python with Real-World Use Cases}.
Medium blog post, 2025.
Available at: \url{https://medium.com/@jagadeesan.ganesh/mastering-llm-ai-agents-building-and-using-ai-agents-in-python-with-real-world-use-cases-c578eb640e35}.

\item \label{G4}
Anthropic Engineering.
\textit{How We Built Our Multi-Agent Research System}.
Technical blog post, 2025.
Available at: \url{https://www.anthropic.com/engineering/multi-agent-research-system}.

\item \label{G5}
SuperAnnotate.
\textit{Multi-Agent LLMs in 2025}.
Blog post, 2025.
Available at: \url{https://www.superannotate.com/blog/multi-agent-llms}.

\item \label{G6}
Solace.
\textit{Long-Term Memory in Agentic AI Systems}.
Blog post, 2025.
Available at: \url{https://solace.com/blog/long-term-memory-agentic-ai-systems/}.

\item \label{G7}
LangChain.
\textit{How and When to Build Multi-Agent Systems}.
Technical blog post, 2025.
Available at: \url{https://blog.langchain.com/how-and-when-to-build-multi-agent-systems/}.

\item \label{G8}
Prompting Guide.
\textit{LLM Agents}.
Research guide article, 2025.
Available at: \url{https://www.promptingguide.ai/research/llm-agents}.

\item \label{G9}
Deepchecks.
\textit{How Multi-Agent LLMs Differ from Traditional LLMs}.
Blog post, 2025.
Available at: \url{https://www.deepchecks.com/how-multi-agent-llms-differ-from-traditional-llms/}.

\item \label{G10}
Bito.
\textit{AI Agent vs LLM: Understanding the Differences}.
Blog post, 2025.
Available at: \url{https://bito.ai/blog/ai-agent-vs-llm/}.

\item \label{G11}
LabelYourData.
\textit{Multi-Agent LLMs: Concepts and Workflow}.
Blog post, 2025.
Available at: \url{https://labelyourdata.com/articles/multi-agent-llm}.

\item \label{G12}
Ruwei Pan, Hongyu Zhang, and Chao Liu.
\textit{CodeCoR: An LLM-Based Self-Reflective Multi-Agent Framework for Code Generation}.
arXiv preprint arXiv:2501.07811, 2025.

\item \label{G13}
freeCodeCamp.org.
\textit{The Open Source LLM Agent Handbook}.
Online handbook, 2025.
Available at: \url{https://www.freecodecamp.org/news/the-open-source-llm-agent-handbook/}.

\item \label{G14}
Microsoft.
\textit{AI Agents in Azure Cosmos DB}.
Microsoft Learn documentation, 2025.
Available at: \url{https://learn.microsoft.com/en-us/azure/cosmos-db/ai-agents}.

\item \label{G15}
Xue-Guang.
\textit{LLMs for Multi-Agent Cooperation}.
Blog post, 2025.
Available at: \url{https://xue-guang.com/post/llm-marl/}.

\item \label{G16}
Zeeshan Rasheed, Malik Abdul Sami, Kai-Kristian Kemell, Muhammad Waseem, Mika Saari, Kari Systä, and Pekka Abrahamsson.
\textit{Codepori: Large-scale system for autonomous software development using multi-agent technology}.
arXiv preprint arXiv:2402.01411, 2024.

\item \label{G17}
Anurag Mishra.
\textit{Future of Coding: Multi-Agent LLM Framework Using LangGraph}.
Medium blog post, 2025.
Available at: \url{https://medium.com/@anuragmishra_27746/future-of-coding-multi-agent-llm-framework-using-langgraph-092da9493663}.

\item \label{G18}
Micheline Bénédicte Moumoula, Serge Lionel Nikiema, Albérick Euraste Djire, Abdoul Kader Kabore, Jacques Klein, and Tegawendé F. Bissyandé.
\textit{Beyond Language Barriers: Multi-Agent Coordination for Multi-Language Code Generation}.
arXiv e-prints, 2025.

\item \label{G19}
JetBrains AI.
\textit{The Future of AI in Software Development}.
Blog post, 2025.
Available at: \url{https://blog.jetbrains.com/ai/2025/07/the-future-of-ai-in-software-development/}.

\item \label{G20}
Tahir Balarabe.
\textit{The Difference Between AI Assistants and AI Agents — and Why It Matters}.
Medium blog post, 2025.
Available at: \url{https://medium.com/@tahirbalarabe2/the-difference-between-ai-assistants-and-ai-agents-and-why-it-matters-03b5ace6055a}.

\item \label{G21}
AugmentCode.
\textit{Why Multi-Agent LLM Systems Fail and How to Fix Them}.
Web guide article, 2025.
Available at: \url{https://www.augmentcode.com/guides/why-multi-agent-llm-systems-fail-and-how-to-fix-them}.

\item \label{G22}
Center for Security and Emerging Technology.
\textit{Multimodality, Tool Use, and Autonomous Agents}.
Online article, 2025.
Available at: \url{https://cset.georgetown.edu/article/multimodality-tool-use-and-autonomous-agents/}.

\item \label{G23}
NVIDIA Developer.
\textit{How Small Language Models Are Key to Scalable Agentic AI}.
Blog post, 2025.
Available at: \url{https://developer.nvidia.com/blog/how-small-language-models-are-key-to-scalable-agentic-ai/}.

\item \label{G24}
AI Multiple.
\textit{8 AI Code Models Benchmarked: LMC-Eval}.
Online article, 2025.
Available at: \url{https://research.aimultiple.com/ai-code-editor/}.

\item \label{G25}
Business Wire.
\textit{Reply Launches Silicon Shoring, an AI Powered Software Delivery Model to Optimise and Automate the Entire Software Development Life Cycle}.
News release, 2025.
Available at: \url{https://www.businesswire.com/news/home/20250515940305/en/Reply-Launches-Silicon-Shoring-an-AI-Powered-Software-Delivery-Model-to-Optimise-and-Automate-the-Entire-Software-Development-Life-Cycle}.

\item \label{G26}
InfoQ.
\textit{LlamaFirewall Adds Agent Protection with Malware Scanning}.
News article, 2025.
Available at: \url{https://www.infoq.com/news/2025/05/llamafirewall-agent-protection/}.

\item \label{G27}
MarkTechPost.
\textit{Democratizing AI: Implementing a Multimodal LLM-Based Multi-Agent System with No-Code Platforms for Business Automation}.
News article, 2025.
Available at: \url{https://www.marktechpost.com/2025/01/10/democratizing-ai-implementing-a-multimodal-llm-based-multi-agent-system-with-no-code-platforms-for-business-automation/}.

\item \label{G28}
Sourena Khanzadeh.
\textit{AgentMesh: A Cooperative Multi-Agent Generative AI Framework for Software Development Automation}.
arXiv preprint arXiv:2507.19902, 2025.

\item \label{G29}
Yiping Jia, Zhen Ming Jiang, Shayan Noei, and Ying Zou.
\textit{MemoCoder: Automated Function Synthesis using LLM-Supported Agents}.
arXiv preprint arXiv:2507.18812, 2025.

\item \label{G30}
IBM.
\textit{AI Agents vs AI Assistants}.
IBM Think article, 2025.
Available at: \url{https://www.ibm.com/think/topics/ai-agents-vs-ai-assistants}.

\item \label{G31}
IBM Research.
\textit{Large Language Models Revolutionized AI. LLM Agents Are What’s Next}.
Research blog post, 2025.
Available at: \url{https://research.ibm.com/blog/what-are-ai-agents-llm}.

\item \label{G32}
Unit 42, Palo Alto Networks.
\textit{AI Agents Are Here. So Are the Threats}.
Security blog post, 2025.
Available at: \url{https://unit42.paloaltonetworks.com/agentic-ai-threats/}.

\item \label{G33}
YouTube.
\textit{Qwen 3 Coder Plus: BEST Agentic Coding LLM! Insanely Powerful, Fast, \& Free! (Open Source)}.
Video, 2025.
Available at: \url{https://www.youtube.com/watch?v=JuU1JQL6S2w}.

\item \label{G34}
YouTube.
\textit{The Best Local AI Agent for Python}.
Video, 2025.
Available at: \url{https://www.youtube.com/watch?v=NIBprn5cEZA}.

\item \label{G35}
Dayu Wang, Jiaye Yang, Weikang Li, Jiahui Liang, and Yang Li.
\textit{Reducing Cognitive Overhead in Tool Use via Multi-Small-Agent Reinforcement Learning}.
arXiv preprint arXiv:2508.08882, 2025.

\item \label{G36}
Pillar Security.
\textit{New Vulnerability in GitHub Copilot and Cursor: How Hackers Can Weaponize Code Agents}.
Security blog post, 2025.
Available at: \url{https://www.pillar.security/blog/new-vulnerability-in-github-copilot-and-cursor-how-hackers-can-weaponize-code-agents}.

\item \label{G37}
Ars Technica.
\textit{Research AI Model Unexpectedly Modified Its Own Code to Extend Runtime}.
News article, 2024.
Available at: \url{https://arstechnica.com/information-technology/2024/08/research-ai-model-unexpectedly-modified-its-own-code-to-extend-runtime/}.

\item \label{G38}
Aofan Liu, Haoxuan Li, Bin Wang, Ao Yang, and Hui Li.
\textit{RA-Gen: A Controllable Code Generation Framework Using ReAct for Multi-Agent Task Execution}.
arXiv preprint arXiv:2510.08665, 2025.

\item \label{G39}
Abhishek Kodati, Foutse Khomh, and Ashkan Sami.
\textit{MAC: Multi-Agent LLM Coder Is All You Need}.
SSRN Working Paper 5887028.

\item \label{G40}
Yuansheng Ni, Songcheng Cai, Xiangchao Chen, Jiarong Liang, Zhiheng Lyu, Jiaqi Deng, Kai Zou, Ping Nie, Fei Yuan, Yue Xiang, et al.
\textit{VisCoder2: Building Multi-Language Visualization Coding Agents}.
arXiv preprint arXiv:2510.23642, 2025.

\end{enumerate}

\printcredits

% ---- References ----
\bibliographystyle{elsarticle-num} % CAS numeric style
\bibliography{References}

\balance
\end{sloppypar}
\end{document}